\documentclass[aps,twocolumn,floatfix,superscriptaddress,nofootinbib]{revtex4-2}

\usepackage{amsmath}
\usepackage{amssymb}
\usepackage{graphicx}
\usepackage[caption=false]{subfig}
\usepackage{dcolumn}
\usepackage{bm}
\usepackage{latexsym}
\usepackage{soul}
\usepackage[colorlinks=true,linkcolor=blue,citecolor=blue]{hyperref}
\usepackage{color}
\usepackage{xcolor}
\usepackage{parskip}

\DeclareFontFamily{U}{mathx}{\hyphenchar\font45}
\DeclareFontShape{U}{mathx}{m}{n}{<-> mathx10}{}
\DeclareSymbolFont{mathx}{U}{mathx}{m}{n}
\DeclareMathAccent{\widebar}{0}{mathx}{"73}

\definecolor{vert}{rgb}{0,0.69,0}
\definecolor{violet}{rgb}{0.69,0,0.69}
\definecolor{orange}{rgb}{0.99,0.46,0}

\newcommand{\be}{\begin{equation}}
\newcommand{\ee}{\end{equation}}

\begin{document}

\title{High-energy acceleration phenomena in extreme radiation-plasma interactions}

\author{J.~C.~Faure}
\email{jeremy.faure@cea.fr}

\author{D.~Tordeux}

\author{L.~Gremillet}
\email{laurent.gremillet@cea.fr}

\affiliation{CEA, DAM, DIF, 91297 Arpajon, France}
\affiliation{Universit\'e Paris-Saclay, CEA, LMCE, 91680 Bruy\`eres-le-Ch\^atel, France}

\author{M.~Lemoine}
\email{lemoine@iap.fr}

\affiliation{Institut d'Astrophysique de Paris, CNRS-Sorbonne Université, F-75014 Paris, France}

\begin{abstract}
We simulate, using a particle-in-cell code, the chain of acceleration processes at work during the Compton-based interaction of a dilute electron-ion plasma with an extreme-intensity, incoherent gamma-ray flux with a photon density several orders of magnitude above the particle density. The plasma electrons are initially accelerated in the radiative flux direction through Compton scattering. In turn, the charge-separation field from the induced current drives forward the plasma ions to near-relativistic speed and accelerates backwards the non-scattered electrons to energies easily exceeding those of the driving photons. The dynamics of those energized electrons is determined by the interplay of electrostatic acceleration, bulk plasma motion, inverse Compton scattering and deflections off the mobile magnetic fluctuations generated by a Weibel-type instability. The latter Fermi-like effect notably gives rise to a forward-directed suprathermal electron tail. We provide simple analytical descriptions for most of those phenomena and examine numerically their sensitivity to the parameters of the problem.
\end{abstract}

\maketitle

\section{Introduction}\label{sec:introduction}

Compact, powerful astrophysical sources are known to exert significant dynamical feedback on their surroundings through the radiation-plasma interactions driven by the copious amounts of non-thermal radiation that they emit. In the Thomson limit, this provides the basis, of course, of the Eddington luminosity argument. Sources powered by an accretion disk, {\it e.g.}, active galactic nuclei, binary systems etc., may thus power outflows through direct radiation pressure or via the Compton rocket effect~\cite{1981MNRAS.196..257S, 1981ApJ...243L.147O, 1982MNRAS.198.1109P, 1991ApJ...383L...7H, 1998MNRAS.300.1047R, 1999MNRAS.305..181B, 2020MNRAS.499.3158C} although, depending on the specific conditions at hand, the same radiative (Compton) interactions may rather result in drag and thus deceleration~\cite{1992ApJ...384..567L, 1996MNRAS.280..781S}. In the extreme case of GRBs, the intense radiative flux associated with the prompt emission phase may itself transmit enough momentum to the interstellar medium to set the plasma in motion with bulk relativistic speeds~\cite{APJ_Madau00, APJ_Thompson00}, thereby modifying the afterglow radiation pattern of the source~\cite{2005ApJ...627..346B, 2007ApJ...671.1877R}. 

Most previous studies have been performed in a fluid or test-particle limit approach, which neglect the microphysical effects associated with the kinetics of the various plasma species. In recent years, the advent of the particle-in-cell (PIC) simulation technique has allowed to explore this regime in a self-consistent way. In particular, Refs.~\cite{APJL_Frederiksen08, PRL_DelGaudio20} have demonstrated that Compton interactions of high-energy photons with electrons can excite electron-accelerating plasma wakes under conditions relevant to the precursors of GRB blast waves. Reference~\cite{JPP_Martinez21} has shown that Compton-driven particle beams in pair plasmas can generate an electromagnetic microturbulence through streaming instabilities. More recently, Ref.~\cite{2022MNRAS.511.3034V} has studied the microphysical consequences of radiative drag in the context of radiation-mediated shock waves, which are expected in the denser parts of astrophysical explosive phenomena~\cite{2010ApJ...725...63B, 2020PhR...866....1L}. Finally, the kinetic physics of Compton backreaction has also been examined in different contexts, in particular radiative relativistic reconnection~\cite{2021MNRAS.508.4532M} and turbulence~\cite{2021ApJ...921...87N}.

Within this context, the present paper addresses the interaction of an extremely intense flux of gamma-ray photons on a fully ionized electron-proton plasma in a general, model-independent setting and provides an in-depth analysis of the various phenomena at play. We tackle this complex problem through large-scale, two-dimensional (2D) PIC simulations, using the \textsc{calder} code~\cite{NucFu_Lefebvre03, theseTordeux}. Our setup complements nicely the previous study of Ref.~\cite{JPP_Martinez21}, which was restricted to electron-positron plasmas. Charge separation between the Compton-scattered electrons and the slowest ions will be shown here to play a crucial role, enabling the acceleration of a fraction of electrons to suprathermal energies, well in excess of those of the driving photon beam. 

The paper is organized as follows. In Sec.~\ref{sec:physcompton}, we explore the physical processes involved in successive stages of the interaction, leading to the formation of several electron sub-populations. This section initially focuses on the region close to the photon front, then addresses the effect of microturbulent electromagnetic fields on the electron momentum distribution as well as the generation of high-energy electron tails propagating both parallel and anti-parallel to the radiative flux. In Sec.~\ref{sec:varparam}, we examine the sensitivity of those phenomena to various initial configuration parameters, by considering, in particular, ions of varying mass, a higher density photon flux, a longitudinally finite photon pulse, and finally different photon energies. We summarize and discuss our results in Sec.~\ref{sec:concl}.

\section{Phenomenology of Compton-driven electron-ion plasmas}
\label{sec:physcompton}

\subsection{Simulation setup} 
\label{subsec:setup}

As already stated, our purpose is to investigate in detail the physical processes at play during the interaction of an intense gamma-ray flux with a tenuous (\emph{i.e.}, effectively collisionless) electron-proton plasma. This scenario is studied through 2D3V (2D in configuration space, 3D in momentum space) particle-in-cell (PIC) simulations in the following configuration.

A monoenergetic, unidirectional photon flux is injected across the left-hand side of the simulation domain and propagates along the longitudinal $\boldsymbol{\hat x}$ axis.
Most of the simulation domain is filled by a fully ionized electron-proton plasma. The transverse photon and plasma density distributions are uniform. The boundary conditions for both particles, photons and field are chosen to be open along $\boldsymbol{\hat x}$ and periodic in the transverse ($\boldsymbol{\hat y}$) direction.

The gamma-ray photons are treated the same way as finite-mass particles, that is, using ``macro-photons'' representing a large number of physical photons. Hence, similarly to what is done for Coulomb binary collisions between plasma particles \cite{PHP_Perez12} or for the Bethe-Heitler pair creation process between photons and ions \cite{PHP_Martinez19}, we have used a Monte-Carlo method \cite{EGS4, theseTordeux} to model the pairwise Compton interactions between the photons and plasma electrons using the Klein-Nishina cross-section \cite{KleinNishina28}. The numerical experiments reported in this work describe both Compton and Coulomb interactions, yet the latter turn out to have a negligible role during the simulated time frame.

The numerical domain is large enough to capture the relevant physics in both directions on the timescale of the numerical experiment. In practice, this means a domain of dimensions $L_x \times L_y = 4\,800 \times 240\,(c/\omega_{\rm p})^2$, discretized into $24\,000 \times 12\,000$ cells with a mesh size $\Delta x = \Delta y = 0.2\,c/\omega_{\rm p}$. Here $c$ denotes the velocity of light and $\omega_{\rm p} \equiv (4\pi n_0 e^2/m_e)^{1/2}$ the electron plasma frequency ($e$ is the elementary charge, $m_e$ the electron mass, and $n_0$ the initial electron number density).  Each (electron or ion) plasma species is initially represented by 20 macro-particles per cell. To reduce their computational cost, most of our simulation use a reduced ion mass of $m_i = 184\,m_e$. The corresponding ion inertial length is thus $c/\omega_{\rm pi} = 13.6\,c/\omega_{\rm p}$.

We adopt the following standard conventions. All throughout, barred spatial and temporal scales are expressed in units of $c/\omega_{\rm p}$ and $\omega_{\rm p}^{-1}$, respectively. Similarly, barred electron (ion) momenta or energies are understood to be normalized quantities expressed in terms of $m_e c$ (resp. $m_i c$) or $m_e c^2$ (resp. $m_ic^2$), respectively.

\begin{figure}
    \centering
    \includegraphics[width=0.45\textwidth]{ 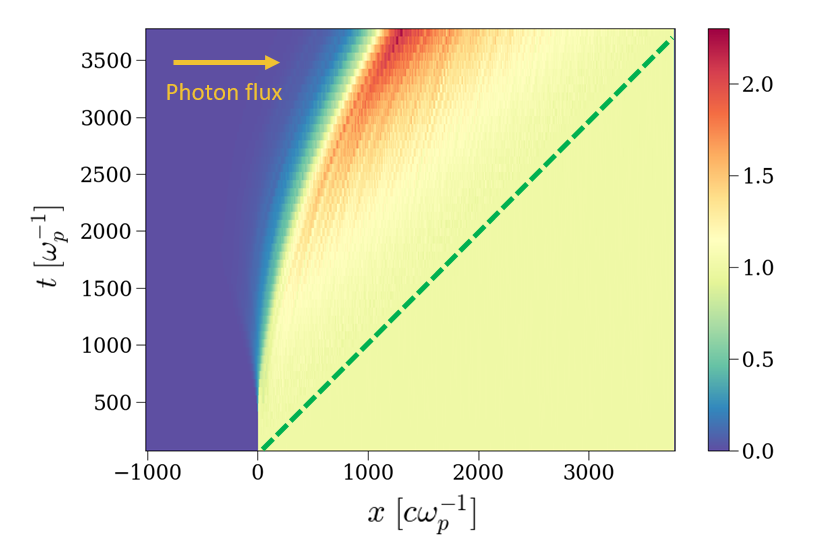}
    \caption{Time evolution of the transversely averaged electron density ($n_e/n_0$). The plasma initially occupies the region $0 \leq x\,\omega_{\rm p}/c \leq 3\,800\,$. The photon flux is impinging from the left-hand side and the green dashed line indicates the trajectory of the radiation front. The simulation parameters are $\bar{\epsilon}_\gamma=4$, $\alpha_\gamma = 1.05\times 10^4$ and $m_i/m_e=184$.}
    \label{fig:np_x_t_E4}
\end{figure}

To avoid introducing a small length scale associated with a sharp-gradient photon front, we inject a photon flux with a longitudinal spatial profile that increases linearly with the distance from its tip (located at $x=0$ at $t=0$).  In the main part of the text, that photon flux profile increases indefinitely towards $-\boldsymbol{\hat x}$, but we discuss the case of a longitudinally finite beam in Sec.~\ref{sec:varparam}. In practice, the \emph{ad hoc} extension of the photon flux does not cause any improper effect because it acts on a slab of plasma of finite longitudinal extent for a finite amount of time. We have run simulations with a constant photon density $n_\gamma$, hence with a sharp front and no diverging flux at the rear. Far from the front of the photon flux, those additional simulations reveal features that are similar to those of the main simulations discussed thereafter, while close to the front, the early stage of the interaction is not as well resolved in time. Everywhere, the distance to the radiative front is defined as $\xi \equiv ct-x$. 

In its initial state, the plasma is uniformly distributed in the simulation box over $3\,800\,c/\omega_{\rm p}$ along $\boldsymbol{\hat{x}}$, leaving a void region of $1\,000\,c/\omega_{\rm p}$ next to the left boundary. The plasma electrons and ions are initialized with equal (isotropic) temperatures $T_e=T_i=100\,\rm eV$ and densities $n_e=n_i=n_0 = 1.1\times 10^{21}\,\rm cm^{-3}$. While the latter density value is chosen with a view to long-term laboratory astrophysics experiments, our simulation results on collective plasma effects can be scaled
to much more dilute astrophysical plasmas ($n_0 \sim 1\,\rm cm^{-3}$) provided the interaction rate per plasma time $\propto\, n_\gamma \omega_{\rm p}^{-1} \,\propto\, (n_\gamma/n_0)\sqrt{n_0}$ is kept constant (at fixed normalized interaction distance $\bar \xi \equiv \omega_{\rm p} \xi/c$ and ion mass).

The incoming photon flux first crosses the void region, then enters the plasma, setting it into motion and compressing it by radiative pressure. This is illustrated in Fig.~\ref{fig:np_x_t_E4} which shows the $x-t$ evolution of the ($y$-averaged) electron plasma density. The right-hand front of the accelerated plasma electrons therefore follows the front of the photon flux, defined by $\xi=0$. The simulation stops before this front can reach the right-hand side of the simulation box. 

\begin{figure*}
    \centering
    \includegraphics[width=0.9\textwidth]{ 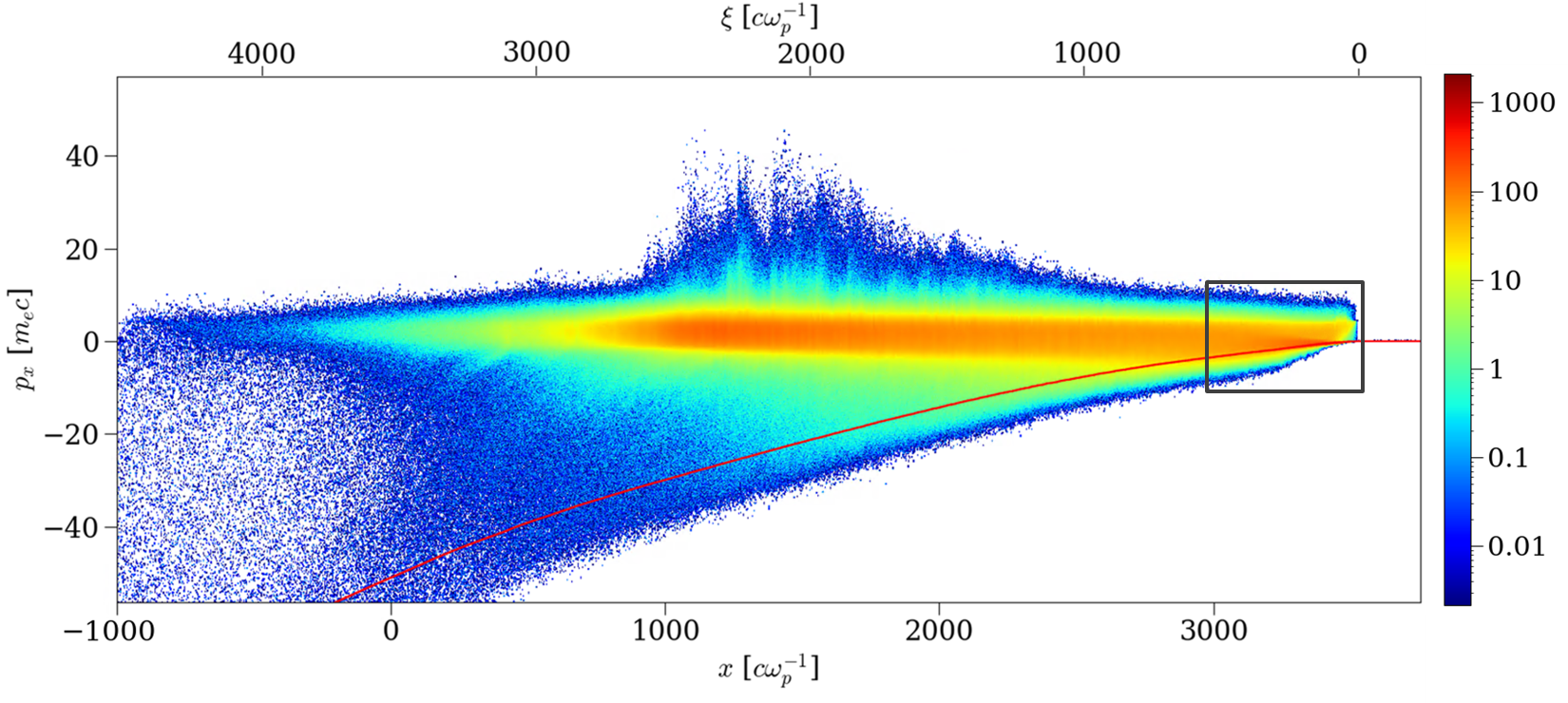}
    \caption{Electron $\xi-p_x$ phase space at time $\bar{t} = 3\,500$, with the corresponding lab-frame $x-$position and  comoving $\xi$-position indicated on the bottom and top axes, respectively. The photon front is then located at $\bar{x} = 3\,500$ ($\bar{\xi} = 0$). The simulation parameters are $\bar{\epsilon}_\gamma = 4$, $\alpha_\gamma = 1.05\times 10^4$ and $m_i/m_e=184$. The black box near the photon front delineates the region where early-time interactions effectively thermalize the plasma bulk. The red line represents the typical trajectory of an electron backward-accelerated by the charge-separation field in the energetic tail.
    }
    \label{fig:qxpx_5e7_E4}
\end{figure*}

At all times, the plasma remains optically thin to the photon flux from back to front. The electron density is indeed such that the Thomson optical depth to interaction, $\tau_{\rm T} \equiv \sigma_{\rm T} n_e  L_x$ ($\sigma_{\rm T}$ denoting the Thomson scattering cross-section), remains very small: $\tau_{\rm T} \simeq 5\times 10^{-5}$. Consequently, the density of the photon flux remains essentially unchanged as it propagates through the plasma. The plasma and electromagnetic field profiles near the radiative front will thus depend on time and space mainly through their combination $\xi$, distance to the photon front.

The main other parameters that characterize the interaction are the photon energy and the ratio of photon to electron densities, respectively set to $\bar{\epsilon}_\gamma \equiv \epsilon_\gamma /m_e c^2 = 4$ and  $n_\gamma/n_0 = \alpha_\gamma\,\bar \xi$ with $\alpha_\gamma = 1.05\times 10^4$ and $\bar \xi_{\rm max} \simeq 3\,800$ in the reference simulation discussed thereafter. The latter choice implies that the mean (normalized) interaction time -- in the Thomson approximation -- for an electron located at $\bar \xi$ is $\bar{t}_{\rm T} \equiv \omega_{\rm p} \left( n_\gamma \sigma_{\rm T}c \right)^{-1} \simeq 8 \times 10^3\,\bar \xi^{-1}$. The condition for at least one Compton interaction is $\bar{t}_{\rm T} \lesssim \bar \xi$, \emph{i.e.}, $\bar \xi \gtrsim 90$. The radiative force will then be felt throughout the plasma crossed by the photon flux. 

Our study being focused on the interplay of Compton scattering and collective plasma effects, we neglect Bremsstrahlung and synchrotron radiation as well as pair production. This is mainly done for the sake of simplicity but can also be justified within a certain parameter range as will be discussed in Sec.~\ref{sec:varparam}.

\subsection{Forward Compton acceleration and charge-separation effects}
\label{subsec:charge_separation}

As the photon flux propagates, it interacts with the plasma electrons through Compton interactions and accordingly sets them into motion. Figure~\ref{fig:qxpx_5e7_E4} provides a global view on the longitudinal $x-p_x$ phase space of the plasma electrons ($p_x$ denoting the $x$-component of the momentum) at an advanced time $\bar{t} = 3\,500$. The many features present in this picture arise through the conjunct effects of Compton scattering, coherent longitudinal electric fields and electromagnetic field fluctuations that result from anisotropies in momentum space. Those various effects are analyzed step by step in the following sub-sections. 

Near the radiative front, the plasma electrons can be split into two main populations: the initially dilute relativistic beam of electrons accelerated to relativistic energies by Compton scattering, and the unscattered plasma bulk. The latter is barely visible in Fig.~\ref{fig:qxpx_5e7_E4} due to its low temperature. Those populations evolve under the action of the two main competing processes, namely, forward Compton scattering due to the incoming photon flux and the backward pull exerted by the ions through a charge-separation field. We quantify those effects in the following.

\begin{figure}
    \centering
    \includegraphics[width=0.45\textwidth]{ 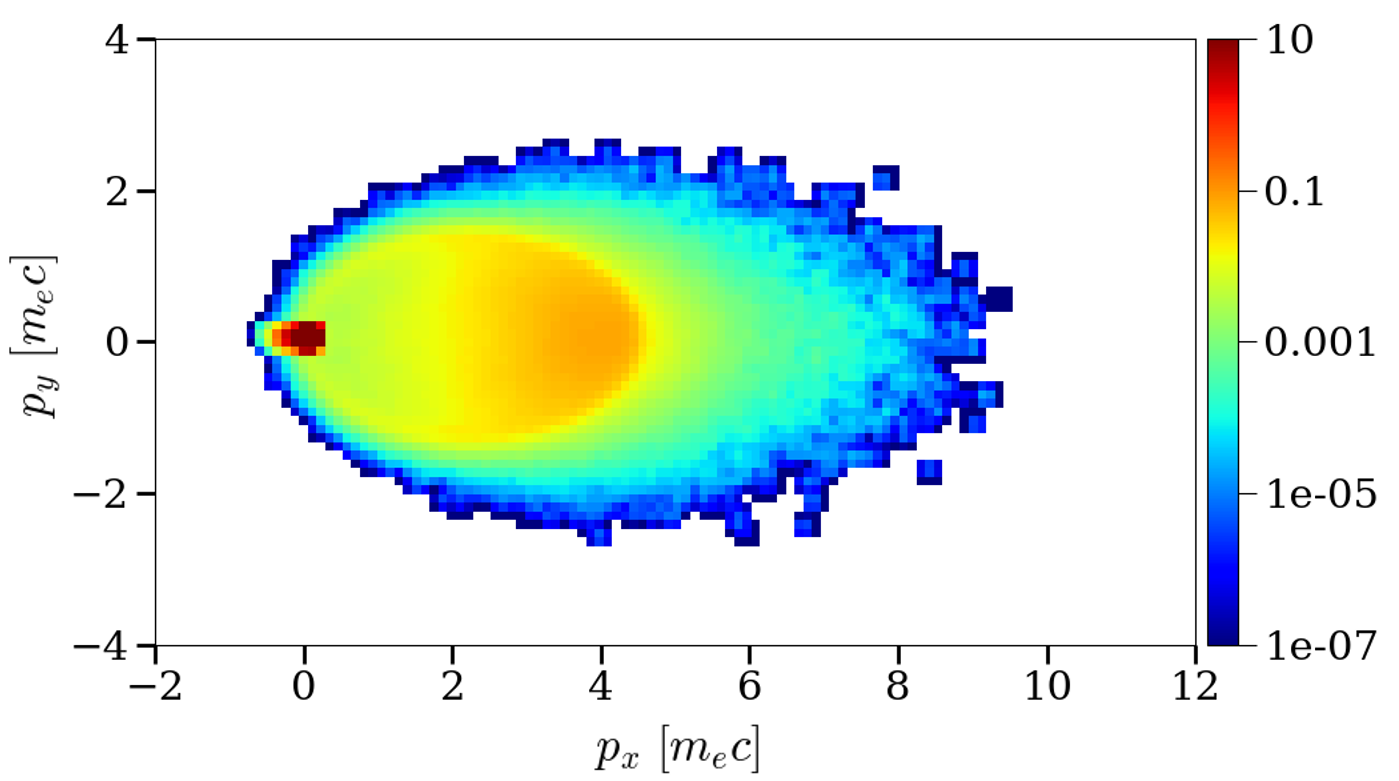}
    \caption{Electron $p_x-p_y$ phase space near the photon front ($0<\bar{\xi}<50$) at $\bar{t}=1\,050$, with $\bar{\epsilon}_\gamma = 4$, $\alpha_\gamma = 1.05\times 10^4$ and $m_i/m_e=184$. The orange region that extends up to $p_x \simeq \epsilon_\gamma/c$ represents the electrons scattered once by the photons, while the green component extending to $\bar{p}_x \simeq 2 \bar{\epsilon}_\gamma$ corresponds to twice scattered electrons. The unscattered dense component is visible in red. The simulation parameters are those of Fig.~\ref{fig:qxpx_5e7_E4}.} 
    \label{fig:pxpy_5e7_avant}
\end{figure}

Cold plasma electrons undergoing Compton scattering are pushed to momenta comparable with those of the photons, since the interaction takes place in the Klein-Nishina regime where the inelasticity approaches a sizable fraction of unity. In detail, the ratio of the mean scattered ($\bar{\epsilon}_{\rm \gamma,s}$) to incoming ($\bar{\epsilon}_\gamma$) photon energies is $\bar{\epsilon}_{\gamma,s}/\bar{\epsilon}_\gamma \simeq (4/3)/(\ln 2\bar{\epsilon}_\gamma + 1/2) \simeq 0.5$ for scattering electrons initially at rest and $\bar{\epsilon}_\gamma = 4$ \cite{APJ-Barbosa82}. 
This corresponds to an average electron energy gain $\Delta \gamma = \bar{\epsilon}_\gamma (1-\bar{\epsilon}_{\gamma,s}/\bar{\epsilon}_\gamma) \simeq 2$. 

Those stochastic Compton interactions exert on the plasma electrons an average radiative force directed toward $+\boldsymbol{\hat x}$ \cite{ApJ_Blumenthal74}. This is clearly seen in Fig.~\ref{fig:pxpy_5e7_avant}, which displays the $p_x-p_y$ momentum distribution of the electrons near the photon front (integrated over $0 < \bar \xi < 50$). There, the scattered electrons form a dilute, relativistic beam moving in the same direction as the incoming photons \cite{PRL_DelGaudio20, JPP_Martinez21}. The thermal unscattered bulk corresponds to the dense red central region near $p \simeq 0$. The density ratio of the electron populations having been scattered once or twice scales approximately with the effective optical depth to interaction.

\begin{figure}
    \centering
    \includegraphics[width=0.45\textwidth]{ 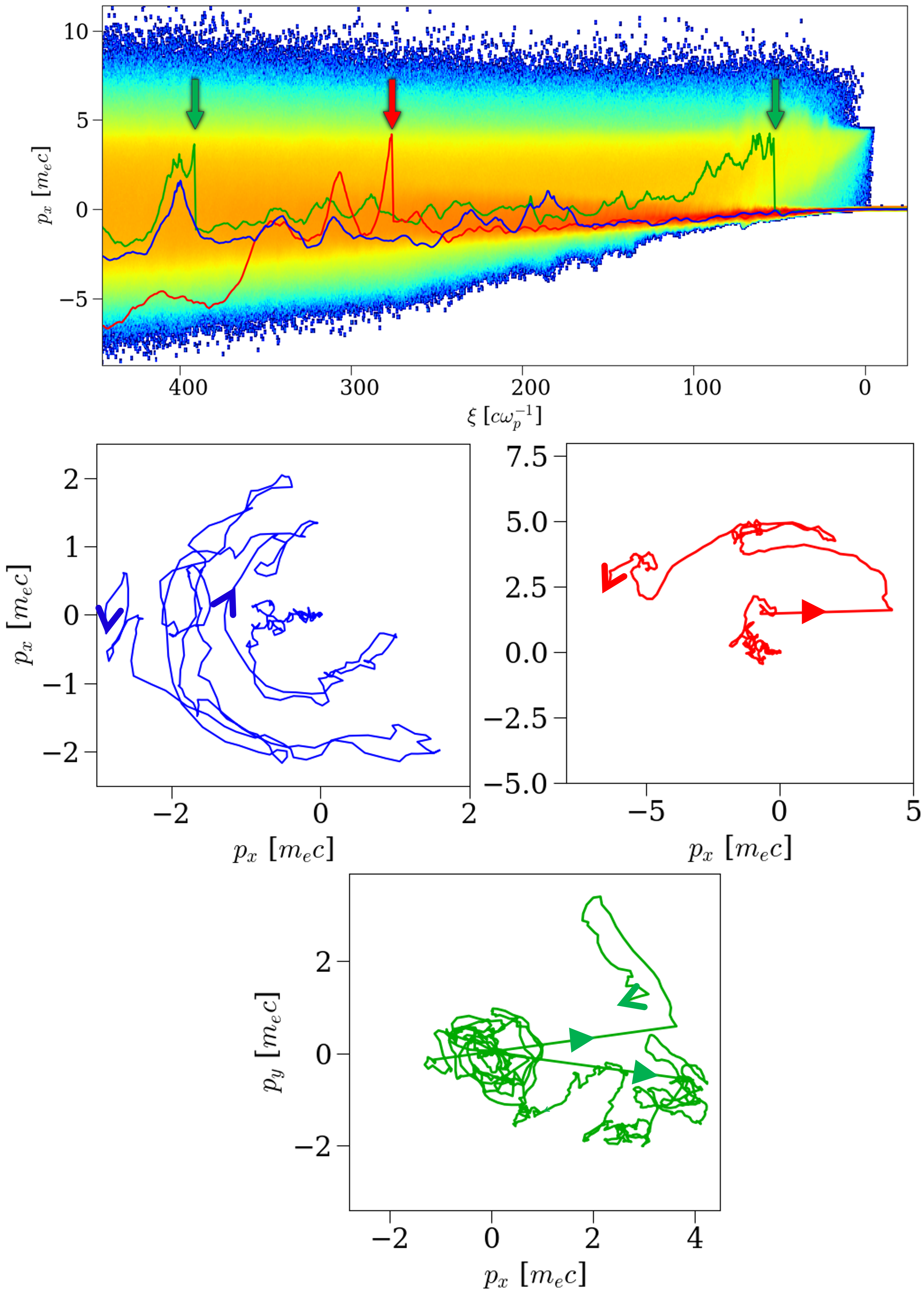}
    \caption{Top: Electron $\xi-p_x$ phase space, near the photon front (as indicated). The blue, red and green lines represent three representative electron trajectories. The colored arrows pinpoint the locations of individual Compton scattering events.
    Bottom: Corresponding trajectories
    in $p_x-p_y$ space. Plain arrows indicate the direction of motion while filled arrows trace the jumps due to Compton scatterings. The simulation parameters are those of Fig.~\ref{fig:qxpx_5e7_E4}.}
    \label{fig:qx-px_avant}
\end{figure}

The charge separation caused by the slower dynamics of the plasma ions gives rise to a coherent longitudinal electric field, $\langle E_x \rangle$. This field accelerates ions forward and electrons backwards, and hence acts to neutralize the electrical current induced by the Compton interactions. In particular, the nonscattered bulk of plasma electrons make up a return current, counterstreaming against the radiative flux. This population appears as a thin, dark red region in the electron $\xi-p_x$ phase shown in the top panel of Fig.~\ref{fig:qx-px_avant}. The typical electron trajectories plotted in the top and bottom panels of Fig.~\ref{fig:qx-px_avant} will be discussed in more detail in Sec.~\ref{subsec:EM_instabilities}. 

The ions being much heavier than the electrons, the electron momentum distribution, $f_e(\mathbf{p})$, can be regarded at a near equilibrium of forces, {\it i.e.}, the mean radiative force $F_{\rm rad}$ exerted on the electrons locally balances the pull due to the $\langle E_x \rangle$ electric component, or 
\be 
e \langle E_x \rangle = \frac{1}{n_e}\,\int \mathrm{d}^2 p\,f_e(\mathbf{p}) F_{\rm rad}(\mathbf{p}) \,,
\label{eq:force-eq}
\ee
where $n_e \equiv \int{\rm d}^2p\, f_e(\mathbf{p})$ is the electron density\footnote{In our 2D3V simulations, owing to the conjunct effect of Compton scattering, acceleration in the $\langle E_x \rangle$ field and deflections in the $B_z$ fluctuations, the electron momenta are mainly distributed in the $p_x-p_y$ plane, and so we consider $f_e(\mathbf{p}) \equiv f_e(p_x,p_y)$ for modeling purposes.}. The mean radiative force $F_{\rm rad}(\mathbf{p})$ exerted on an electron of Lorentz factor $\gamma$ and momentum $\mathbf{p} \equiv \gamma \boldsymbol{\beta} m_e c$ is calculated by integrating over scattering angles the variation in photon momentum, properly weighted by the interaction rate using the full Compton cross-section~\cite{ApJ_Blumenthal74, APJ_Madau00, APJ_Zampieri03}. For a monoenergetic, unidirectional ($x$-directed) photon flux interacting with the plasma electrons, this can be written in the following compact form~\cite{ApJ_Blumenthal74}
\be
F_{\rm rad}(\mathbf{p}) = \sigma_{\rm T} n_\gamma \epsilon_\gamma G(\eta) (1-\beta_x) \left[1-\gamma^2 \beta_x \frac{1-\beta_x}{1+\eta} \right] \,, 
\label{eq:Frad_dist_elec}
\ee
where $\eta= \bar{\epsilon}_\gamma \gamma (1- \beta_x)$ and 
\begin{align}
G(\eta) &= \frac{3(1+\eta)}{4\eta^3} \left[ \frac{\eta^2-2\eta-3}{2\eta} \ln (1+2\eta) \right. \nonumber \\
&\left. + \frac{-10\eta^4+51\eta^3+93\eta^2+51\eta+9}{3(1+2\eta)^3}\right] \,.
\label{eq:Frad_F}
\end{align}
It therefore directly follows that $\langle E_x \rangle$ scales approximately linearly with the photon density $n_\gamma$ at fixed other parameters. The function $G(\eta)$ incorporates the Klein-Nishina correction \cite{KleinNishina28} to momentum transfer when \mbox{$\eta \gg 1$}. 

\begin{figure}
    \centering
    \includegraphics[width=0.45\textwidth]{ 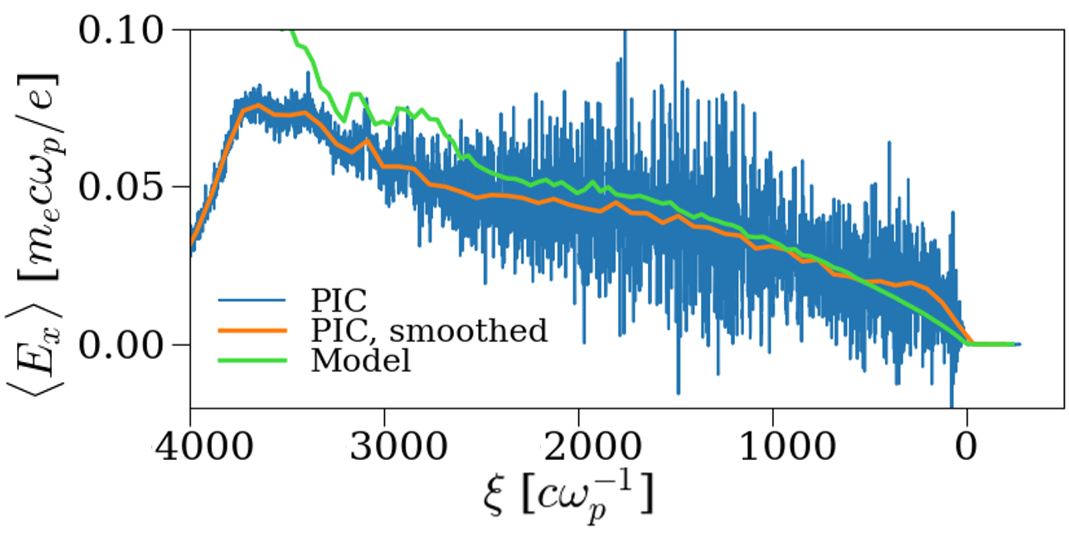}
    \caption{Blue: Longitudinal profile of the transversely averaged $\langle E_x \rangle$ field extracted at $\bar{t} = 3\,500$ from the same PIC simulation as in Fig.~\ref{fig:qxpx_5e7_E4}. Orange: same quantity but smoothed over 100 grid points along $\xi$. Green: Approximate $\langle E_x \rangle$ field as given by Eq.~\eqref{eq:force-eq} and fitting the local electron distribution function to a bi-Maxwellian distribution. See text for details.}
    \label{fig:tracethE_5e7}
\end{figure}

Extracting $f_e(\mathbf{p})$ from our simulation, we can integrate Eq.~\eqref{eq:force-eq} to infer $\langle E_x \rangle$. To simplify the calculation, we use a local drifting bi-Maxwellian fit to $f_e(\mathbf{p})$. In detail, for each position $\xi$, we define a $p_x-p_y$ momentum distribution of the form $f_{\rm M}( \mathbf{\bar{p}}) \,\propto\, e^{\frac{-(\bar{p}_x-\bar{p}_d)^2}{2\widebar{T}_x}-\frac{\bar{p}_y^2}{2 \widebar{T}_y}}$, whose longitudinal and transverse temperatures $\widebar{T}_x$ and $\widebar{T}_y$, and drifting momentum $\bar{p}_d$ are adjusted so as to reproduce the local first moments $\langle \bar{p}_x \rangle$, $\langle \bar{p}_x^2\rangle$ and $\langle \bar{p}_y^2\rangle$ of the simulated distribution $f_e(\bar{\xi},\mathbf{\bar{p}})$ (averaged along $y$). With such an approximation, the computation of Eq.~\eqref{eq:force-eq} is greatly eased and the inferred longitudinal profile of $\langle E_x \rangle$ matches well with that measured in the simulation, see Fig.~\ref{fig:tracethE_5e7}. In particular, one can observe that $\langle E_x \rangle$ increases about linearly with the distance to the photon front, as expected given our choice of $n_\gamma \,\propto\, \bar{\xi}$.

As detailed in Appendix~\ref{app:Ex}, we have obtained approximate expressions for $\langle E_x \rangle$ in the case where the plasma electron momenta are distributed according to a drifting relativistic Maxwell-J\"uttner function, $f_e(\mathbf{\bar {p}}) \,\propto \,e^{-\mu' \gamma_d(\gamma - \beta_d \bar{p}_x)}$, with $\mu' \equiv 1/\widebar{T}'$ as the inverse proper temperature and $\beta_d$ as the drift velocity.
For relativistically hot electrons ($\mu' \ll 1$) and in the Klein-Nishina limit ($\eta \gg 1$), one obtains
\begin{align}
    &\frac{e \langle E_x \rangle}{m_e c \omega_{\rm p}} \simeq \frac{3}{8} \sigma_{\rm T} \frac{c}{\omega_{\rm p}} \frac{\alpha_\gamma n_0 \bar{\xi}}{\bar{\epsilon}_\gamma} \Bigg\{\left(\frac{\mu' \bar{\epsilon}_\gamma}{\gamma_d} - \beta_d \right) \nonumber \\
    & \times \left[ \ln \left(\frac{\bar{\epsilon}_\gamma}{\mu'} \sqrt{\frac{1-\beta_d}{1+\beta_d}} \right) - \gamma_{\rm E} - \frac{5}{6}\right] + 1- \beta_d \Bigg\} \,.
    \label{eq:Ex_edf_KN_approx}
\end{align}
Assuming $1/\mu' = \mathcal{O}(\bar{\epsilon}_\gamma)$ and $\gamma_d \sim 1$, the electric field is thus predicted to scale as $\langle E_x \rangle \,\propto\, \bar{\epsilon}_\gamma^{-1}$ as the photon energy and, concomitantly, the mean plasma energy increase. The above formula also depicts how the electric field weakens when the electron bulk velocity approaches $c$ ($\gamma_d > 1$), as it eventually happens here because of the simultaneous acceleration of the plasma ions (see below).

For completeness, we have also approximated $\langle E_x \rangle$ in the Thomson limit ($\eta \ll 1$) when $\widebar{T} \equiv 1/\mu  = \mathcal{O}(\bar{\epsilon}_\gamma) \ll 1$. Assuming negligible bulk motion, one finds
\be
\frac{e\langle E_x \rangle}{m_e c \omega_{\rm p}} \simeq \sigma_{\rm T} \frac{c}{\omega_{\rm p}} n_0 \alpha_\gamma \bar{\xi} \bar{\epsilon}_\gamma \left(1 + 2\bar{T} -\frac{11}{5}\bar{\epsilon}_\gamma \right) \,. 
\label{eq:Ex_edf_Thomson_approx}
\ee
To leading order, the electric field then varies with the photon energy as $\langle E_x \rangle \,\propto\, \bar{\epsilon}_\gamma$, in stark contrast with the Klein-Nishina regime.

\subsection{Dynamics of the ions}
\label{sec:iondyn}

\begin{figure}
    \centering
    \includegraphics[width=0.45\textwidth]{ 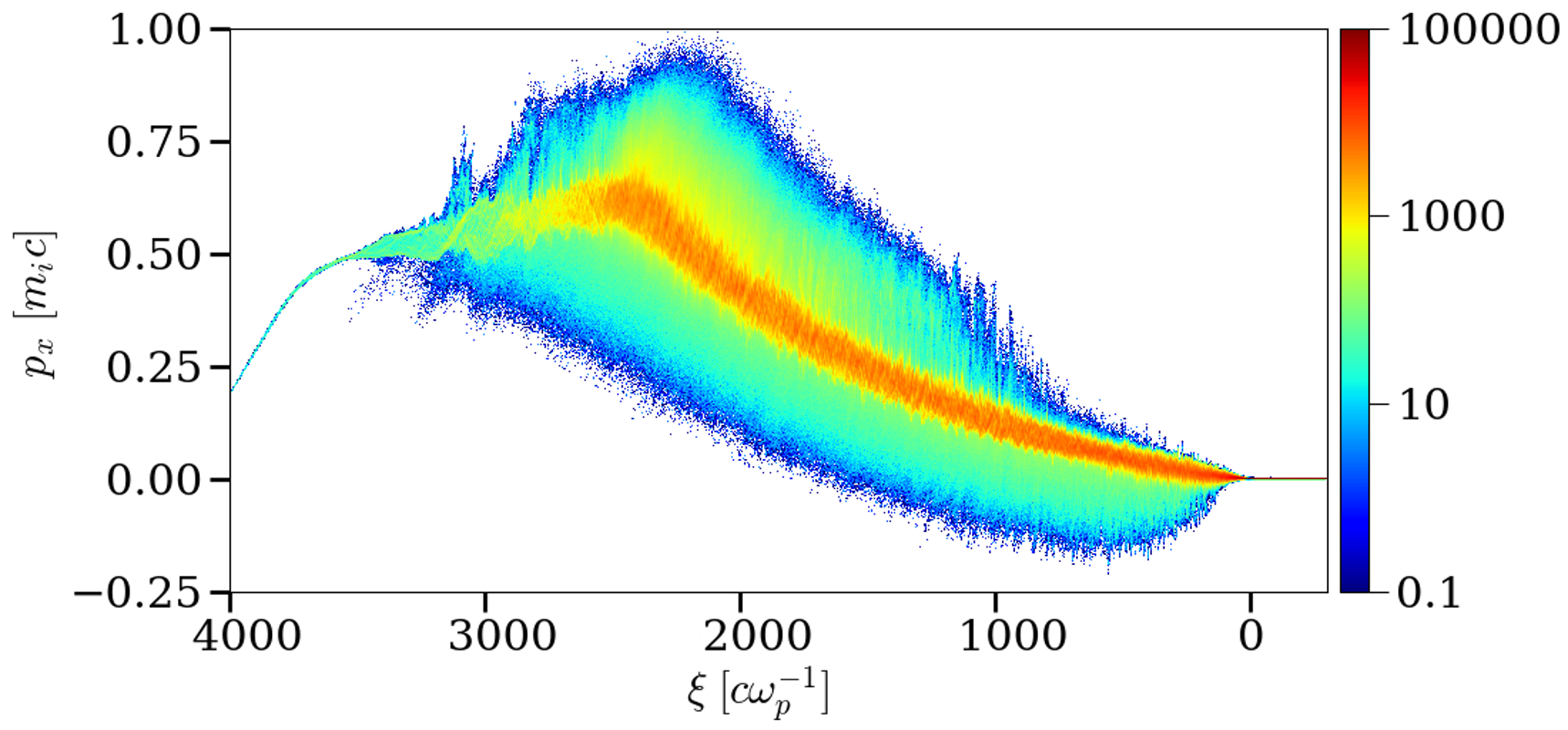}
    \caption{Ion $\xi-p_x$ phase space at time $\bar{t} = 3\,500$ from the same simulation as in Fig.~\ref{fig:qxpx_5e7_E4}. The overall profile in mean momentum $p_x$ is shaped by the electric drag, as discussed in the text.}
    \label{fig:qxpx_5e7_ion}
\end{figure}

We now address the dynamics of the plasma ions acted upon by the coherent electric field. Figure~\ref{fig:qxpx_5e7_ion} shows their $\xi-p_x$ phase space at time $t=3\,500\,\omega_{\rm p}^{-1}$. The bulk ions see their $x$-momentum increase to $p_x/m_ic \simeq 0.6$ over a $\sim 2\,500\,c/\omega_{\rm p}$ distance to the radiative front, while the fastest ions reach $p_x/m_ic \simeq 1$. This drag effect and the accompanying ion compression (by a factor of $\sim 2$ at $\bar{t} \simeq 3\,500$) are also manifest in Fig.~\ref{fig:np_x_t_E4}. 

The bulk ion dynamics can be modeled using a cold-fluid approximation, assuming negligible transverse motion and that all relevant quantities (ion density $n_i$, Lorentz factor $\gamma_i$, normalized $x$-velocity $\beta_{ix}$ and $x$-momentum $\bar{p}_{ix} \equiv \gamma_i \beta_{ix}$) depend on $\bar \xi$ only. The conservation of the longitudinal ion current can thus be expressed as 
\be
\frac{\mathrm{d}}{\mathrm{d} \bar \xi}\,\left[(1-\beta_{ix}) n_i \right] = 0 \,. 
\label{eq:ion_continuity}
\ee
The evolution of the ion momentum obeys
\be 
\left(1-\beta_{ix} \right) \frac{\mathrm{d} \bar{p}_{ix}}{\mathrm{d} \bar \xi} = \frac{e \langle E_x \rangle}{m_i c \omega_{\rm p}} \,,
\label{eq:dui_dt_1}
\ee
which can be recast as
\be
\frac{\mathrm{d}}{\mathrm{d} \bar \xi} \left(\bar{p}_{ix}-\gamma_i \right) = \frac{e\langle E_x\rangle}{m_i c \omega_{\rm p}} \,.
\label{eq:dui_dt_2}
\ee
A straightforward integration then yields
\be
\bar{p}_{ix} - \gamma_i = \frac{e}{m_i c \omega_{\rm p}} \int d \bar \xi \, \langle E_x \rangle - 1 \,. 
\label{eq:dui_dt_3}
\ee
Based on Eq.~\eqref{eq:Ex_edf_KN_approx}, we now consider that the electric field is proportional to the photon density, $\langle E_x \rangle = E_0\alpha_\gamma \bar \xi$.
This leads to
\be
\bar{p}_{ix} - \gamma_i = A \bar{\xi}^2 - 1 \,,
\ee
where $A \equiv eE_0 \alpha_\gamma/(2m_ic \omega_{\rm p})$. Since $\gamma_i = \sqrt{1+\bar{p}_{ix}^2}$, one finally obtains
\be
\bar{p}_{ix} =  A\bar{\xi}^2 \frac{(2-A\bar{\xi}^2)}{2(1-A \bar{\xi}^2)} \,.
\label{eq:ionspeed}
\ee
In the simulation, upon fitting the area below the $\langle E_x \rangle$ curve in Fig.~\ref{fig:tracethE_5e7} over over $0 \le \bar{\xi} \le 2\,000$, we extract $E_0 \simeq 2.8\times 10^{-9}\,m_e c \omega_{\rm p}/e$, giving $A \simeq 8.1\times 10^{-8}$. The above formula then predicts $\bar{p}_{ix}(\bar{\xi}=2000) \simeq 0.40$, in very good agreement with what is observed in Fig.~\ref{fig:qxpx_5e7_ion}.

Initially, $A \bar{\xi}^2 \ll 1$ so that $\bar{p}_{ix} \simeq A\bar{\xi}^2$. When $\bar{\xi}$ approaches $A^{-1/2}$, the ion momentum increases to infinity, yet in that case, the hypothesis of $\langle E_x \rangle \,\propto\, \bar \xi$ no longer holds due to
the spatial dependence of $\beta_d$
in Eq.~\eqref{eq:Ex_edf_KN_approx_5}, not taken into account in the integrand of Eq.~\eqref{eq:dui_dt_3}.

To conclude this part, we assess the accuracy of our analytical estimate of $\langle E_x \rangle$ from the knowledge of the ion velocity profile, assuming $\beta_d \simeq \beta_{ix}$. At $\bar \xi = 1\,000$, we measure in the simulation $\beta_{ix} \simeq 0.1$ and $\mu' \simeq 0.6$; Equation~\eqref{eq:Ex_edf_KN_approx_5} then predicts $\langle E_x \rangle \simeq 0.02\,m_e c\omega_{\rm p}/e$, in agreement with Fig.~\ref{fig:tracethE_5e7}. At $\bar \xi = 2\,000$, the bulk speed is larger ($\beta_{ix} \simeq 0.41$) and $\mu'$ has increased to $\simeq 0.72$, leading to an excessively low electric field, $\langle E_x \rangle \simeq 0.008\,m_e c\omega_{\rm p}/e$ , starkly contrasting with the PIC value, $\langle E_x \rangle \simeq 0.047\,m_e c\omega_{\rm p}/e$.
This inconsistency, which is resolved by using a bi-Maxwellian fit (as done in Fig.~\ref{fig:tracethE_5e7}), may stem from our questionable modeling of the plasma electrons with a relativistic drifting Maxwell-J\"uttner distribution. Also, 
it may arise from the fact that the argument of the $\ln$ term in Eq.~\eqref{eq:Ex_edf_KN_approx_5} is
only marginally large ($\frac{\bar{\epsilon}_\gamma}{\mu'}\sqrt{\frac{1-\beta_d}{1+\beta_d}} \simeq 3.6$), unlike what is assumed in the Klein-Nishina limit. 

\subsection{Electromagnetic instabilities due to mixing electron populations}
\label{subsec:EM_instabilities}

\begin{figure}
    \centering
    \includegraphics[width=0.47\textwidth]{ 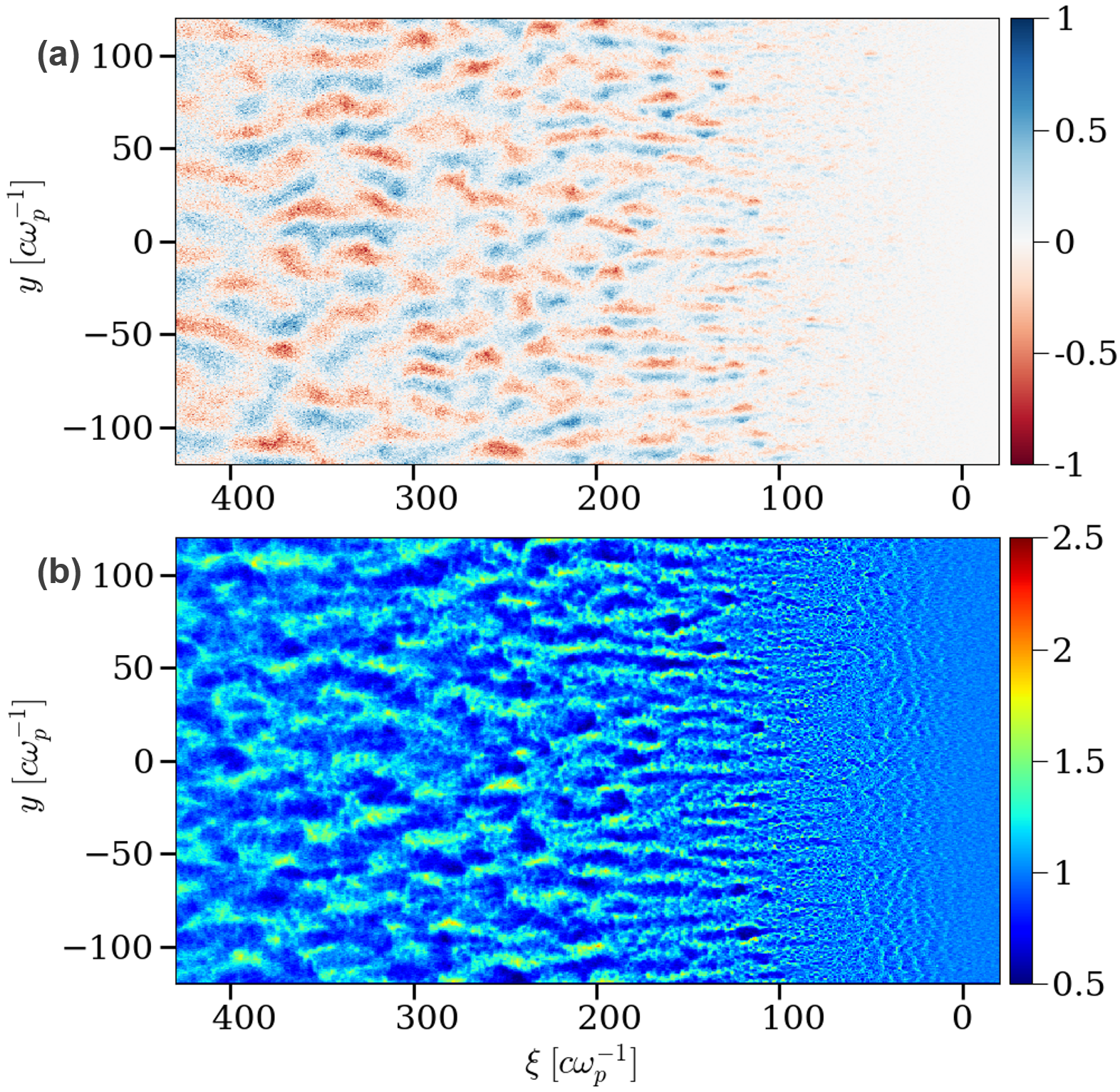}
    \caption{(a) Out-of-plane magnetic field ($eB_z/m_e \omega_{\rm p} c$) and (b) electron density ($n_e/n_0$) distributions near the photon front. The simulation parameters are those of Fig.~\ref{fig:qxpx_5e7_E4}.}
    \label{fig:Bz_avant}
\end{figure}

As discussed earlier and illustrated in Fig.~\ref{fig:qx-px_avant}, the electron distribution near the photon front comprises two main groups, one forward-scattered by the photons, one backward-accelerated by the charge-separation electric field. This counterstreaming configuration is notoriously prone to a number of electromagnetic microinstabilities~\cite{PRL_Bret08, *PRE_Bret10, *PoP_Bret10}. Here, we notably observe the growth of the Weibel-type, magnetic current filamentation instability (CFI)~\citep{PRL_Weibel59, PoF_Fried59, PoF_Davidson72, PRL_Lee73, PRE_Califano97, PoP_Honda00, PoP_Silva02, PoP_Bret07, AA_Achterberg07a, *AA_Achterberg07b, PRE_Pelletier19, PRE_Bresci22}. Figure~\ref{fig:Bz_avant}(a) indeed reveals the formation of transverse magnetic modulations of typical wavelength $\bar{\lambda}_\perp \simeq 13$ and amplitude $\langle 2 B_z^2\rangle^{1/2} \simeq 0.28 \,m_e c \omega_{\rm p}/e$ at $\bar{\xi} \gtrsim 200$. This instability, which first amplifies current and magnetic modulations, eventually generates dense ($\delta n_e /n_e \simeq 0.5$) electron filaments surrounded by intense transverse magnetic fields in its nonlinear stage [Fig.~\ref{fig:Bz_avant}(b)]. For $\bar{\xi} \gtrsim 100$, those magnetic fields are found to carry an energy much exceeding that associated with the $E_x$ and $E_y$ electric fields, which demonstrates the dominance of the CFI over competing electrostatic streaming instabilities \cite{PRE_Bret10, *PoP_Bret10}.

In turn, the CFI-driven fields affect the electrons that were Compton scattered in the forward direction. Those are repeatedly deflected in the magnetic structures that form in the course of the instability and they can be effectively reflected to negative $x$-momenta. 

Such a sequence of events can be seen in Fig.~\ref{fig:qx-px_avant} (bottom panel), where are plotted the trajectories of three macro-electrons near the photon front in $p_x-p_y$ space. Consider for instance the green macro-electron: it first undergoes a strong Compton scattering event (indicated by a filled arrow) at $\bar{\xi} \simeq 60$, bringing it to $\bar{p}_x \simeq 4$, and then follows a random walk across the still relatively weak CFI fields ($\langle 2 B_z^2\rangle ^{1/2} \simeq 0.15\,m_e \omega_{\rm p} c/e$, $\bar{\lambda}_\perp \simeq 9$), while decelerating in the coherent electric field. After returning to the plasma bulk ($\bar{p} \lesssim 1$), a second Compton scattering occurs at $\bar{\xi} \simeq 400$, where the magnetic filaments have grown stronger and wider ($\langle 2\, B_z^2\rangle^{1/2} \simeq 0.32\,m_e c \omega_{\rm p}/e$, $\bar{\lambda}_\perp \simeq 25$) [Fig.~\ref{fig:Bz_avant}(a)]. The electron Larmor radius, $r_L = pc/eB \simeq 8\,c/\omega_{\rm p}$, then becomes comparable with the filament size, and hence the electron may next experience quasi-magnetostatic interactions. This is actually what happens just afterwards: the electron gyrates anticlockwise for some time while keeping its absolute momentum approximately constant, and before suffering a final deflection in the opposite direction.

In comparison, the red macro-electron experiences its first Compton scattering only at $\bar \xi \simeq 280$, where the CFI fields have already reached a large amplitude, and so its subsequent trajectory proceeds in a series of circular segments characteristic of quasi-magnetostatic interactions.

The blue macro-electron experiences a slightly different history: it does not undergo Compton scattering, therefore retains a smaller momentum (note the difference in scale between the three plots), rotates around magnetic filaments near $p_x=0$ and at the same time suffers progressive acceleration to $p_x<0$ values in the longitudinal charge-separation field. 

Through angular scattering, the microturbulence endows the plasma electrons with an effective collision frequency, which redistributes the momentum gained along $+\boldsymbol{\hat{x}}$ by Compton scattering or along $-\boldsymbol{\hat{x}}$ by the coherent electric field. Altogether, this leads to plasma heating, which is seen as the broadening of the central red region in Fig.~\ref{fig:qx-px_avant} (top panel). According to that figure, the distinction between the various electron populations appears to fade progressively at distances $\bar \xi \gtrsim 200$ away from the radiative front. This value also corresponds roughly to the distance beyond which electrons have been Compton scattered at least once. The mean number of Compton scatterings that an electron (initially at rest) suffers as it moves from $\bar \xi$ to $\bar \xi+{\rm d} \bar{\xi}$ is indeed ${\rm d}N_{\rm C} = n_\gamma(\bar{\xi}) \sigma_{\rm C}(c/\omega_{\rm p}) {\rm d} \bar{\xi}$, with $n_\gamma(\bar{\xi}) = \alpha_\gamma n_0\,\bar{\xi}$ ($\sigma_{\rm C}$ is the Compton cross-section \cite{KleinNishina28} and we recall that $\alpha_\gamma =1.05 \times 10^4$). Consequently, at $\bar \xi$, an electron has been scattered $N_{\rm C}(\bar{\xi}) \simeq 0.5 \alpha_\gamma n_0 \sigma_{\rm C} \bar{\xi}^2 \,c/\omega_{\rm p}$ times, so that the distance beyond which an electron has been scattered once is $\bar{\xi} \simeq \left(2 \omega_{\rm p}/\alpha_\gamma n_0 \sigma_{\rm C}c\right)^{1/2} \simeq 280$.

\begin{figure}
    \centering
    \includegraphics[width=0.4\textwidth]{ 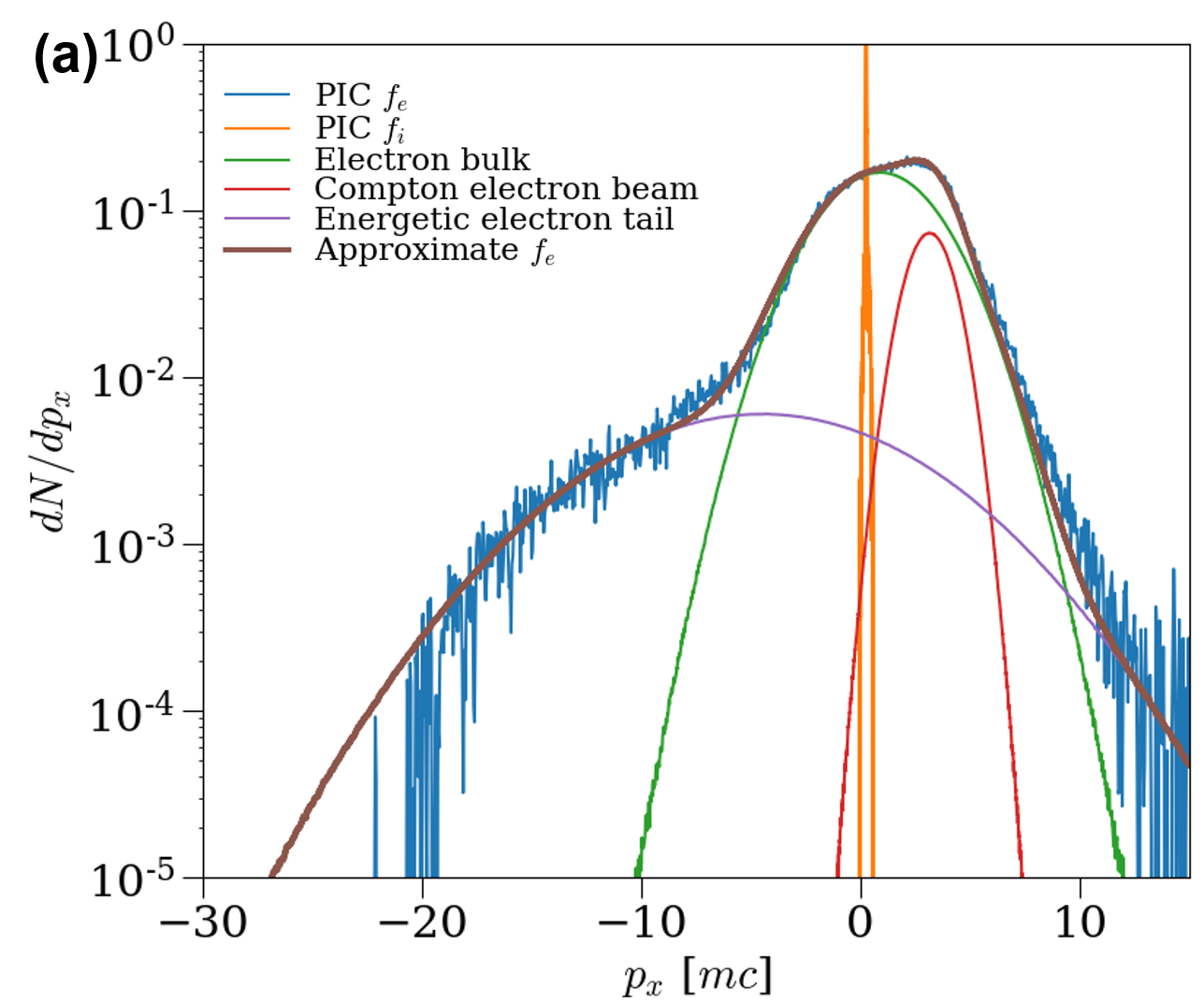}
    \includegraphics[width=0.4\textwidth]{ 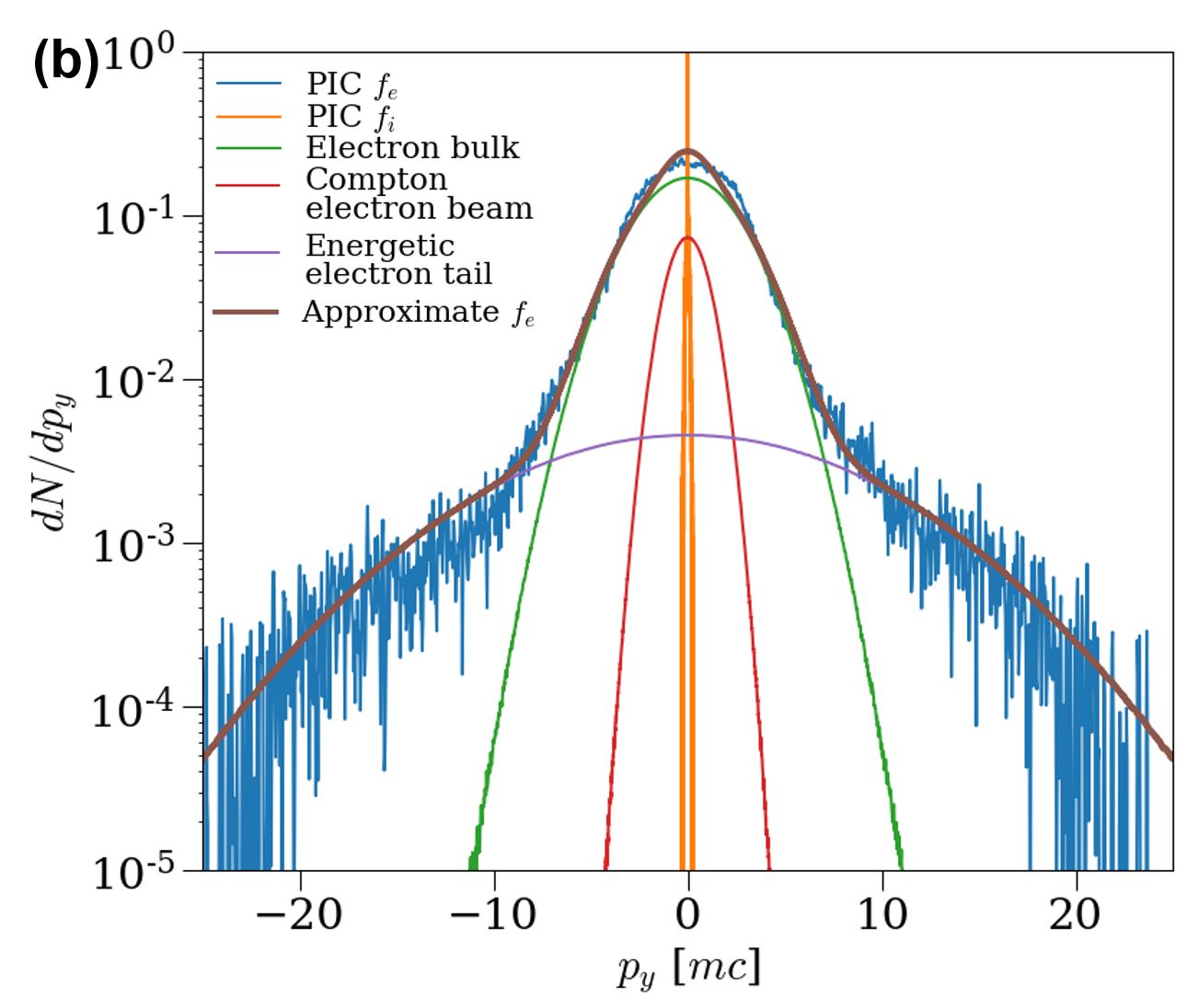}
    \caption{(a) Longitudinal and (b) transverse momentum distributions of the plasma (blue) electrons and (orange) ions, as extracted at $\bar{\xi}=1\,500$ from the same simulation as in Fig.~\ref{fig:qxpx_5e7_E4}, and averaged over the transverse dimension. The electron (ion) momenta are normalized to $m_ec$ (resp. $m_i c$).
    The brown curves closely matching the simulated electron distribution in both panels is the sum of the three drifting bi-Maxwellian components plotted in green, red and purple. See Eq.~\eqref{eq:edf_fit} and text for details.}
    \label{fig:px_5e7_xi1500}
\end{figure}

Further away from the photon front, the coexistence of plasma groups with different drift velocities and momentum dispersions keeps feeding the development of a CFI-type instability. Those various populations are illustrated in Fig.~\ref{fig:px_5e7_xi1500}, which plots the electron and ion (longitudinal and transverse) momentum distributions at $\bar \xi = 1\,500$. The ion distribution assumes the form of a narrow peak, shifted at small yet positive $p_x$. The electron distribution exhibits a broader and more complex shape, which can be roughly approximated as the sum of three drifting bi-Maxwellian components, namely,
\be
f_e(\mathbf{\bar{p}}) \simeq \sum_{j=1}^3 f_{\rm Mj} (\mathbf{\bar{p}}) \,,
\label{eq:edf_fit}
\ee
where $f_{\rm Mj}(\mathbf{\bar{p}}) = \frac{n_j}{2\pi \widebar{T}_j} e^{-\frac{(\bar{p}_x-\bar{p}_{dj})^2}{2\widebar{T}_{jx}} - \frac{\bar{p}_y^2}{2 \widebar{T}_{jy}}}$. Here $n_j$ represents the number density, $\bar{p}_{dj}$ the mean $x$-momentum and $\widebar{T}_{jx}$ ($\widebar{T}_{jy}$) the longitudinal (resp. transverse) temperature for the $j$th component. Those components represent, respectively, the electron plasma bulk ($n/n_e \simeq 0.79$, $\bar{p}_d \simeq +0.88$, $\widebar{T}_x = \widebar{T}_y  \simeq 6.3$), the (backward-propagating) energetic electron tail ($n/n_e \simeq 0.07$, $\bar{p}_d \simeq -4.5$, $\widebar{T}_x \simeq 39$, $\widebar{T}_y \simeq 68$), and a colder population dubbed the Compton electron beam ($n/n_e \simeq 0.14$, $\bar{p}_d \simeq +3.15$, $\widebar{T}_x = \widebar{T}_y \simeq 1.0$), accounting for the increase in $f_e(p_x)$ near the photon energy. The good accuracy of this model distribution is illustrated in Fig.~\ref{fig:px_5e7_xi1500}. Due to the continuing action of the charge-separation field and Compton scattering, this nonequilibrium distribution is maintained at all times far from the photon front, and so is the CFI.

\begin{figure}
    \centering
    \includegraphics[width=0.45\textwidth]{ 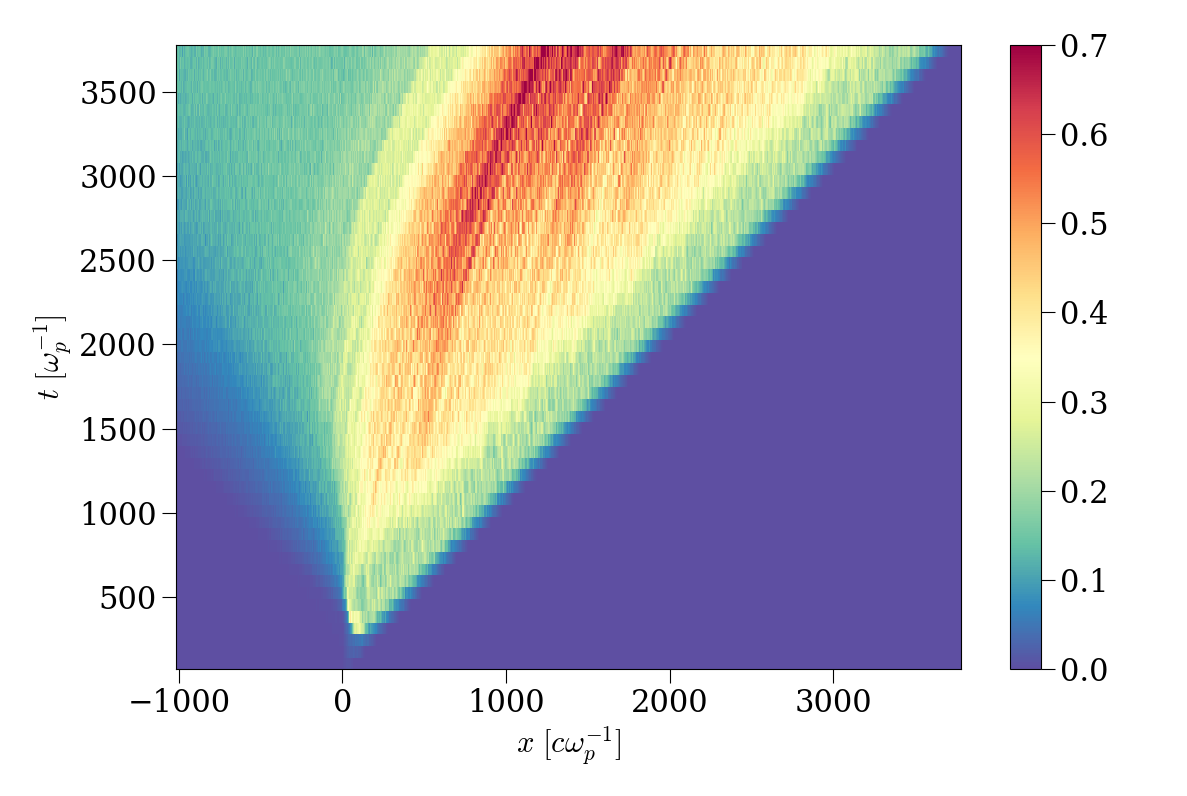}
    \caption{Evolution of the root-mean-square magnetic field $\langle B_z^2 \rangle^{1/2}$ (averaged over the transverse dimension and normalized to $m_e c \omega_{\rm p}/e$) as a function of longitudinal position and time. The moving CFI-driven magnetic structures appear as the coherent streaks developing at late times. Same simulation parameters as in Fig.~\ref{fig:qxpx_5e7_E4}.}
    \label{fig:Bz_x_t_E4}
\end{figure}

One should notice that the magnetic structures resulting from the CFI at large $\xi$ are not stationary in the lab frame. This is best seen in Fig.~\ref{fig:Bz_x_t_E4}, which displays the $x-t$ evolution of the root-mean-square of $B_z$, averaged over the transverse direction. The massive ions dictate the bulk motion of the microturbulence so that the filaments, and the magnetic fields surrounding them, move in the lab frame at a speed close to that of the ions while slowly deforming. The ion frame here coincides with the Weibel frame defined in Ref.~\onlinecite{PRE_Pelletier19}, which moves at $\beta_{\rm w} = \langle E_y^2\rangle^{1/2}/\langle B_z^2\rangle^{1/2}$ and in which the microturbulence can be regarded as essentially magnetostatic. We have checked that $\beta_{\rm w} \simeq \beta_{ix}$ in the region of well-formed filaments.

To identify which electron and ion populations are mainly responsible for the CFI activity far from the front, we have performed simplified, fully periodic simulations that consider the approximate electron distribution defined by Eq.~\eqref{eq:edf_fit} and a drifting ion population (either ballistic or responsive to the fields) that initially ensures current neutralization.
We have found that a necessary condition for reproducing qualitatively the CFI development seen in the radiation-plasma simulation is the presence of both the electron bulk and Compton beam components (\emph{i.e.}, the green and red curves in Fig.~\ref{fig:px_5e7_xi1500}).
Said otherwise, the backward-propagating electron tail and the drifting plasma ions weakly contribute to the moving magnetic structures arising at large distances.

\subsection{Formation of a backward-propagating suprathermal electron tail}
\label{subsec:suprathermal_tail}

Most of the electrons mix in the plasma bulk through the processes discussed earlier, {\it i.e.}, Compton processes and electric drag in conjunction with scattering off magnetic filaments. Meanwhile, a small fraction of the plasma electrons that evade Compton scattering for hundreds of $c/\omega_{\rm p}$ are progressively accelerated backwards to relativistic energies that eventually exceed those of the driving photons. As they do so, the Compton cross-section falls off because of Klein-Nishina corrections~\cite{KleinNishina28, 1970RvMP...42..237B}. More quantitatively, the cross-section behaves as
$\sigma_{\rm C} \simeq (3/4)\sigma_{\rm T} \, \ln (2\eta)/(2\eta)$ at $\eta \gg 1$, where $\eta = 2\bar{\epsilon}_\gamma \gamma$ represents the (normalized) photon energy seen by an electron moving towards $-\boldsymbol{\hat x}$. Those electrons then behave as ``runaway'' electrons, as the probability of being scattered back into the bulk decreases as their energy increases in the charge-separation field. This picture must, however, be nuanced. In the standard scenario, the friction force exerted on a runaway electron is due to collisions with bulk electrons, and hence is a decreasing function of its velocity. Here, the friction force corresponds to the radiative force, Eq.~\eqref{eq:Frad_dist_elec}, which happens to slowly (logarithmically) increase with the electron energy, $F_{\rm rad} \simeq (3/8)\sigma_{\rm T} n_\gamma m_e c^2 \left(\ln 2\eta -5/6 \right)/\bar{\epsilon}_\gamma$ in the $\gamma \gg \bar{\epsilon}_\gamma$ limit. Yet, as will now be shown, this does not preclude the electron distribution from developing a suprathermal tail.

In the absence of Compton scattering, the electron energy evolves as
\be 
\frac{{\mathrm{d}} \gamma^{(0)}}{\mathrm{d}\bar{\xi}} = -\frac{e \langle E_x \rangle }{m_e c \omega_{\rm p}} \frac{\beta_x^{(0)}}{(1-\beta_x^{(0)})} \,.
\label{eq:px_path1} 
\ee
The superscript $^{(0)}$ serves to mark such scatter-free trajectories, as illustrated by the red curve in Fig.~\ref{fig:qxpx_5e7_E4}, which considers $\beta_x^{(0)} = -0.6$. Electrons accelerated to $\gamma^{(0)} \gg 1$ by $\langle E_x \rangle$ have $\beta_x^{(0)} \simeq -1$, yet angular scattering off magnetic filaments may entail $\beta_x^{(0)} \gtrsim -1$. We thus retain the dependence on $\beta_x^{(0)}$ in the following, but consider it as a fixed parameter for simplicity.

To delimit the region of phase space that gives rise to runaway electrons, it then proves convenient to define the ($\xi$- and $p$-dependent) energy variation $\Delta \gamma_r(\xi,\gamma)$, which corresponds to the energy gained by the electron as it drifts against $\langle E_x \rangle$ on a Compton scattering length scale $\xi_{\rm C} = (n_\gamma \sigma_{\rm C})^{-1}$. 
Inserting the various dependencies (including $\langle E_x \rangle = E_0 \alpha_\gamma \overline{\xi}$) and considering the limit $\bar{\xi} \gg (\omega_p/2\sigma_C n_e \alpha_\gamma c)^{1/2}$ that is relevant here, we obtain
\begin{align}
    \Delta \gamma_r(\xi,\,p) & \simeq \frac{eE_0 \alpha_\gamma \bar{\xi}}{m_e \omega_p c} \frac{\vert \beta_x^{(0)} \vert}{1-\beta_x^{(0)}} \bar{\xi}_C \nonumber \\
    &\simeq \frac{8}{3}\frac{e E_0 }{m_e c^2 n_0 \sigma_{\rm T}} \frac{\bar{\epsilon}_\gamma \gamma^{(0)}}{\ln \left(2 \eta^{(0)} \right)} \vert \beta_x^{(0)} \vert \,,
\label{eq:def_pr2}
\end{align}
where $\eta^{(0)} = \bar{\epsilon}_\gamma \gamma^{(0)} (1-\beta_x^{(0)})$. In the present case ($\bar{\epsilon}_\gamma = 4$, $\alpha_\gamma =1.05\times 10^4$), we measure $E_0 \simeq 2.8\times 10^{-9}\,m_e c\omega_{\rm p}/e$ 
in the simulation over $0 < \bar{\xi} < 2\,000$ (see Fig.~\ref{fig:tracethE_5e7}). 
Taking $\gamma^{(0)} \simeq 4$ and $\beta_x^{(0)} \simeq -1$ for the electrons deflected by the near-front magnetic fields, we find $\Delta \gamma_r \simeq 2.4$, a value comparable with the photon energy.

Using the estimate of $E_0$ that follows from Eq.~\eqref{eq:Ex_edf_KN_approx}, $\Delta \gamma_r$ can be further approximated to 
\begin{align}
    &\Delta \gamma_r \simeq \frac{\gamma^{(0)} \vert \beta_x^{(0)}\vert}{\ln 2 \eta^{(0)} } \Bigg\{\left(\frac{\mu' \bar{\epsilon}_\gamma}{\gamma_d} - \beta_d \right) \nonumber \\
    & \times \left[ \ln \left(\frac{\bar{\epsilon}_\gamma}{\mu'} \sqrt{\frac{1-\beta_d}{1+\beta_d}} \right) - \gamma_{\rm E} - \frac{5}{6}\right] + 1- \beta_d \Bigg\} \,. 
\end{align}
For $\mu' \bar{\epsilon}_\gamma = \mathcal{O}(1)$ and $\vert \beta^{(0)} \vert \simeq 1$, and provided $\beta_d$ is weak enough that the term inside brackets weakly varies, one
finds the scaling $\Delta \gamma_r \,\propto\, \gamma^{(0)}/\ln(\ldots \gamma^{(0)})$. 

\begin{figure}
    \centering
    \includegraphics[width=0.45\textwidth]{ 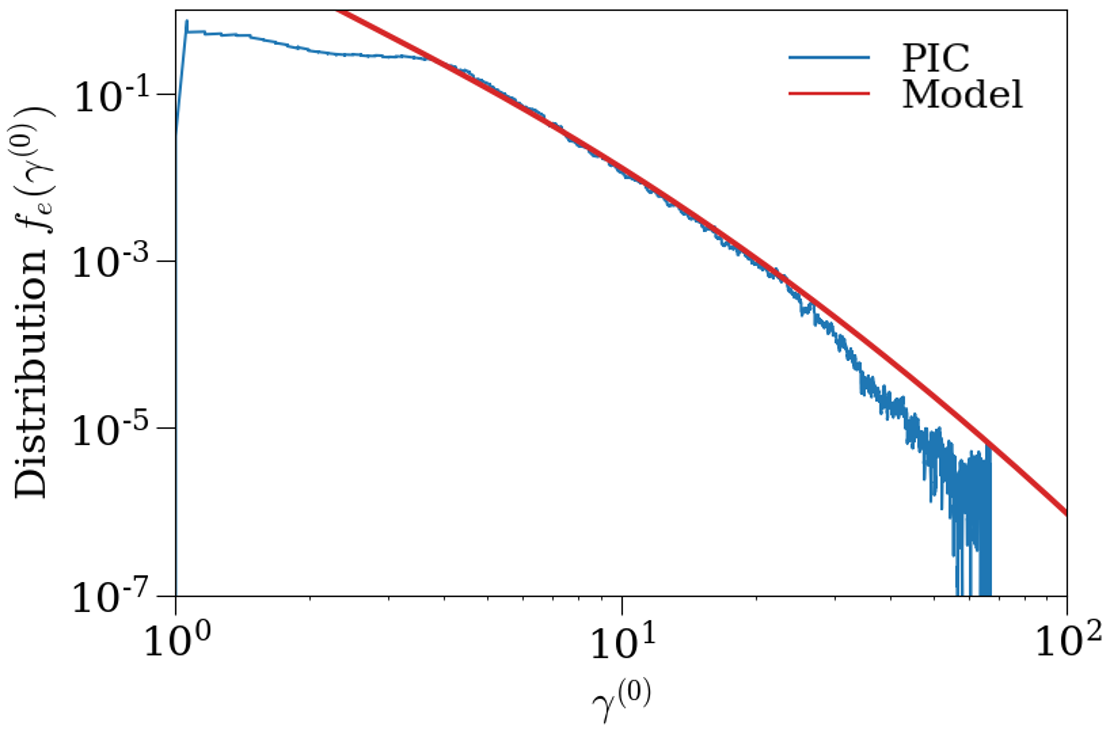}
    \caption{Electron density distribution $f_e(\gamma^{(0)})$ measured in the simulation along an unscattered trajectory $\gamma^{(0)}(\xi)$ as discussed in the text (blue), compared with its analytical approximation, Eq.~\eqref{eq:solfx} (red). The parameter $\beta_x^{(0)}$ is fixed here to $-0.6$, and the amplitude is rescaled to match the simulation data. }
    \label{fig:diagth_ne}
\end{figure}

One may then derive an analytical approximation to the electron distribution function in that high-energy tail with $p_x<0$. This distribution function is evaluated along the unscattered trajectory $\gamma^{(0)}(\xi)$ [see Eq.~\eqref{eq:px_path1}], and is therefore a function of $\xi$ alone. The number density of particles around momentum $\gamma^{(0)}$ changes through Compton scattering processes: particles of energy $\gamma^{(0)}$ are lost through Compton down-scattering but particles of larger energy can come to populate that range through similar down-scattering processes. The latter can be safely neglected, however, because both the cross-section and the distribution function decrease fast with increasing momenta. In this approximation, the distribution function $f_e^{(0)}(\xi)$ evolves as
\be 
    \left(1-\beta_x^{(0)}\right)\frac{\mathrm{d} f_e^{(0)}}{\mathrm{d}\xi} = -n_\gamma\sigma_{\rm C}\left(1-\beta_x^{(0)}\right) f_e^{(0)} \,.\ee
\label{eq:evolfx0}
The prefactor $1-\beta_x^{(0)}$ on the left-hand side comes from the change of variables, from $t$ and $x$ to $\xi$ only, while that on the right-hand side stems from the interaction probability (noting that the cross-section $\sigma_{\rm C}$ is evaluated in the electron rest frame). This equation can be rewritten in a compact way by expressing $f_e^{(0)}$ as a function of $\gamma^{(0)}$,
\be
    \frac{\mathrm{d} f_e^{(0)}}{\mathrm{d} \gamma^{(0)}} = -\frac{f_e^{(0)}}{\Delta \gamma_r} \,.
\label{eq:evolfx0_2}
\ee
Using Eq.~\eqref{eq:def_pr2}, the above can be readily integrated as
\begin{align}
    f_e^{(0)}(\gamma^{(0)}) \,\propto\, \exp \bigg[ &-\frac{3}{16}\frac{m_e c^2 n_0 \sigma_{\rm T}  }{e E_0 \bar\epsilon_\gamma \vert \beta_x^{(0)}\vert} \nonumber \\
    &\times \ln^2\left[2\bar\epsilon_\gamma \gamma^{(0)}\left(1-\beta_x^{(0)}\right)\right] \bigg] \,.
\label{eq:solfx}
\end{align}
In Fig.~\ref{fig:diagth_ne}, we compare this approximation to the distribution function $f_e(\gamma_e^{(0)})$ as computed in the simulation along the unscattered trajectory $\gamma^{(0)}(\xi)$ indicated by a red line in Fig.~\ref{fig:qxpx_5e7_E4} (using the simulated electric field). Setting $\beta_x^{(0)}=-0.6$  allows us to capture fairly well the measured distribution, which takes on the shape of a powerlaw with a slowly-changing slope. The cutoff around $\gamma \simeq 70$ corresponds simply to the point beyond which no electron is accelerated due to the finite length of the numerical domain.

The emergence of the high-energy tail described by Eq.~\eqref{eq:solfx} implies that its energy width be at least comparable with that of the electron bulk distribution, $\widebar{T} = \mathcal{O}(\bar{\epsilon}_\gamma)$.
Now, Eq.~\eqref{eq:solfx} can be recast as
\be
    f_e^{(0)}(\gamma^{(0)}) \,\propto \, \exp \bigg( -\frac{\gamma^{(0)}}{2\Delta \gamma_r} \ln 2\eta^{(0)} \bigg) \,,
\label{eq:solfx_Gr}
\ee
meaning that the typical energy width of $f^{(0)}$ is $2 \Delta \gamma_r/\ln 2\eta^{(0)} = \mathcal{O}(\bar{\epsilon}_\gamma)$ for $\gamma^{(0)} \simeq \bar{\epsilon}_\gamma$ and $\beta_d$ not too large. It then appears that the formation of a suprathermal tail is an intrinsic phenomenon in a plasma irradiated at high photon energies ($\bar \epsilon_\gamma > 1$).

By contrast, as will be numerically demonstrated in Sec.~\ref{subsec:photon_energies}, no suprathermal tail arises when $\bar{\epsilon}_\gamma \ll 1$. Repeating the previous calculation in the Thomson limit indeed yields
\be
    \Delta \gamma_r = \frac{eE_0}{m_e c \omega_{\rm p} n_0 \sigma_{\rm T}} \frac{\vert \beta_x^{(0)}\vert}{1-\beta_x^{(0)}} \,,
    \label{eq:def_pr_thomson_1}
\ee
Upon using Eq.~\eqref{eq:Ex_edf_Thomson_approx} to evaluate $E_0$, one obtains the approximation
\be
    \Delta \gamma_r \simeq \bar{\epsilon}_\gamma \frac{\vert \beta_x^{(0)}\vert}{1-\beta_x^{(0)}} \,.
    \label{eq:def_pr_thomson_2}
\ee
to leading order in $\bar{T} \ll 1$ and $\bar{\epsilon}_\gamma \ll 1$. In the case of a plasma heated to $\bar{T} \simeq \bar{\epsilon}_\gamma$, one has $\beta_x^{(0)} \simeq -\bar{\epsilon}_\gamma^{1/2}$, hence the scaling $\Delta \gamma_r \,\propto \, \bar{\epsilon}_\gamma^{3/2} \ll \bar{\epsilon}_\gamma$ which precludes formation of a suprathermal tail.

The actual plasma dynamics, though, is more complicated than described by our model. Inspection of Fig.~\ref{fig:qxpx_5e7_E4} indeed reveals that the phase-space distribution of electrons, at large $\xi$ and large negative $p_x$, is not a sole function of $\xi$ or $\gamma^{(0)}$, but actually depends on both quantities, a behavior ascribed to the departure of $\langle E_x\rangle$ from a linear scaling in $\xi$. This deviation, visible in Fig.~\ref{fig:tracethE_5e7}, can itself be attributed to the differential motion of the ions between the front and the rear of the plasma slab. Where ions are set into motion, the restoring electric force slightly diminishes, which explains the deviation between the PIC and theoretical distributions at high $\gamma$ in Fig.~\ref{fig:diagth_ne}. This scenario will be confirmed in Sec.~\ref{subsec:immobile_ions} by the analysis of a scenario with infinitely massive ions.

\subsection{Scattering of high-energy electrons off magnetic filaments}
\label{subsec:magnetic_scattering}

Far from the photon front, the moderate-energy electrons of the plasma bulk are efficiently scattered by the magnetic filaments, even though they repeatedly suffer Compton interactions bringing them to energies $\simeq \epsilon_\gamma$. Figure~\ref{fig:Bz_x_t_E4} indicates that the characteristic magnetic field strength in those filaments is $B_z \simeq 0.5\, m_e \omega_{\rm p} c/e $. When expressed as the ratio $\epsilon_B$ of magnetic energy density to the kinetic energy density ($\simeq \gamma_e^2 n_e m_e c^2$) carried by the Compton-scattered electron component moving at Lorentz factor $\gamma_e \simeq \bar{\epsilon}_\gamma$), this corresponds to $\epsilon_B \simeq 1\,\%$, which is a generic value observed at saturation of the CFI (see {\it e.g.}~\cite{PRE_Bresci22} and references therein). As the typical transverse scale of those magnetic filaments is $\bar{\lambda}_\perp \simeq 10$, only electrons with Lorentz factor $\gamma \gtrsim 5$ have a gyroradius larger than the filament size. Such a scattering off a microturbulence moving at about the ion speed tends to reduce the mean longitudinal speed of the electron bulk, as indicated above for the energetic $p_x< 0$ component. 

\begin{figure}
    \centering
    \includegraphics[width=0.45\textwidth]{ 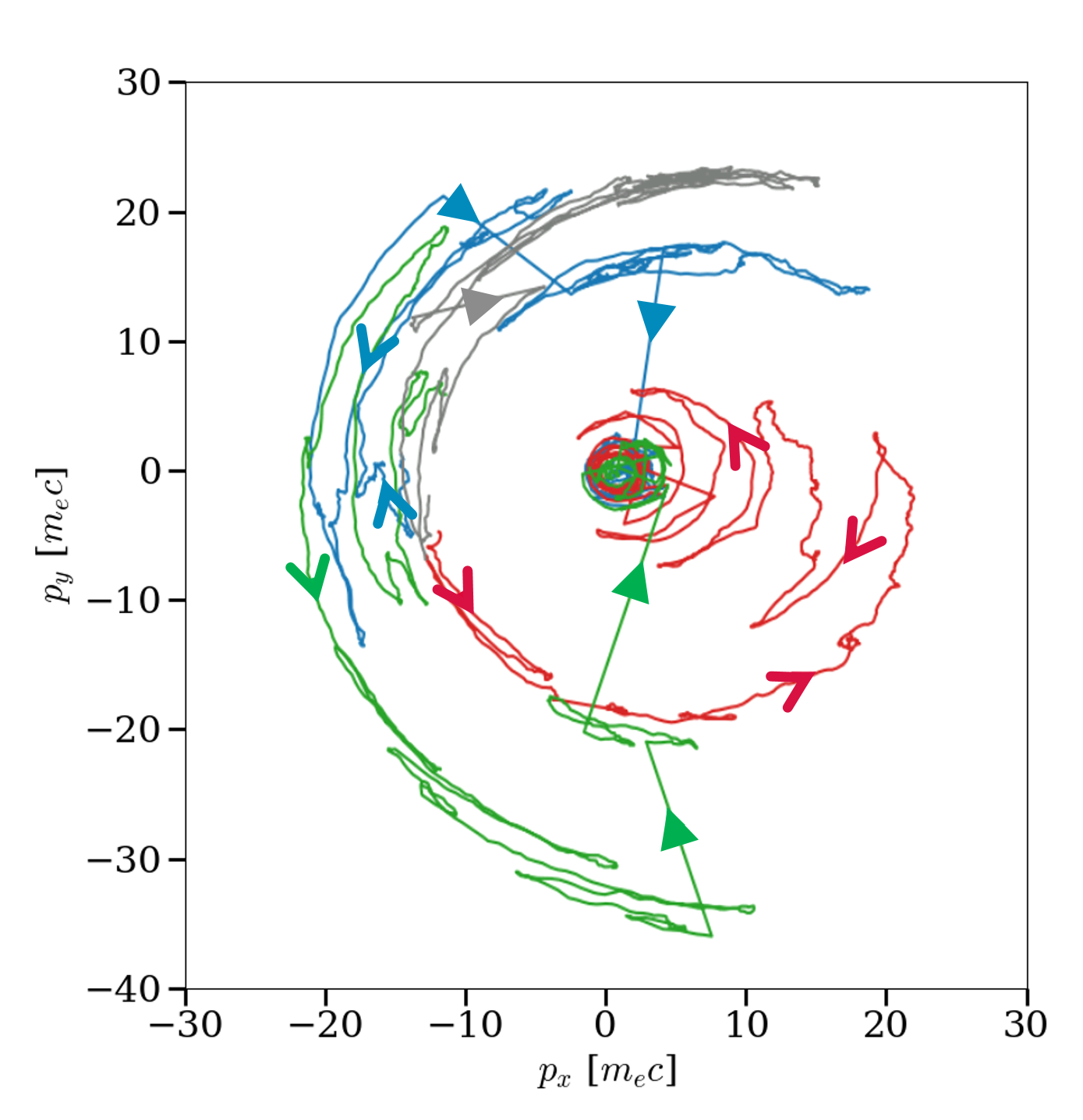}
    \caption{Three representative electron trajectories from the simulation with $\bar{\epsilon}_\gamma = 4$, $\alpha_\gamma =1.05\times 10^4$ and $m_i/m_e=184$. Plain and filled arrows respectively indicate direction of motion and Compton scattering events.}
    \label{fig:pxpy_5e7_reflection_arrow}
\end{figure}

We plot in Fig.~\ref{fig:pxpy_5e7_reflection_arrow} the trajectories of three energetic macro-electrons. As before, plain arrows indicate the direction of motion while filled arrows indicate Compton scattering events. Trajectories in the plasma bulk cannot be distinguished on that figure, as a result of the large scattering frequency in magnetic filaments. However, the central part of the figure, where the trajectories of the energetic electrons end up, gives a rough visualization of the phase space that they occupy.

Electrons in the energetic tail oscillate along a vaguely circular trajectory in phase space due to the effect of alternating positive and negative magnetic field. This is nicely illustrated in the same Fig.~\ref{fig:pxpy_5e7_reflection_arrow} by the green and blue trajectories. As they oscillate, they progressively gain more energy due to the coherent electric field (for instance, the green macro-electron goes from $\bar p_x \simeq -12$ to $\bar p_x \lesssim -20$), forming the energetic tail. Compton scattering gives rise to sudden changes in the momentum of electrons, most of the time severely decreasing their energy. This provides evidence of inverse Compton scattering, which occurs because the electron energy is much higher than the photon energy \cite{APJ-Barbosa82}. After a few scatterings, electrons are then brought back to the low-energy plasma bulk. 

Some of the high-energy electrons can reach positive longitudinal momentum through scattering off the filaments before being brought back toward the low-energy bulk, either through inverse Compton scattering, or under the influence of the charge-separation field which acts to lower positive $p_x$ momenta. One such electron trajectory is represented in red in Fig.~\ref{fig:pxpy_5e7_reflection_arrow}: it is first deflected from $\bar p_x\simeq -12$ to $\bar p_x\simeq +22$, then slowly moves toward low $\bar{p}_x>0$ values just as it oscillates through scatterings in the turbulence.

It is worthwhile to note that the first phase of that process is accompanied by stochastic acceleration in the way envisaged by E.~Fermi~\cite{1949PhRv...75.1169F}, since the magnetized structures behave as moving magnetic mirrors. If the momentum transfer takes place in the direction of those moving structures, {\it i.e.}, towards $+\boldsymbol{x}$ here, then particles gain energy. In detail, if $p_<$ represents the (relativistic) electron momentum before interaction and $p_>$ after, then 
\be
p_> \simeq p_<\,\frac{1-\beta_{ix} \cos\theta_<}{1-\beta_{ix} \cos\theta_>} \,,
\label{eq:stochastic}
\ee
where $\theta_<$ and $\theta_>$ represent, respectively, the angles between the momentum before and after interaction with the (longitudinal) $x$-axis. For a mirror-type reflection from $\cos\theta_<=-1$ to $\cos\theta_>=+1$, as is the case for the red electron in Fig.~\ref{fig:pxpy_5e7_reflection_arrow}, and an observed value of $\beta_{ix} \simeq 0.37$ at the corresponding location, this gives $p_>/p_< \simeq 2.1$, which corresponds well with what is observed.

Hence, a fraction of the energetic tail that has been pulled to large $p_x<0$ values can be reflected with a net energy gain into the forward direction. This population is visible in Fig.~\ref{fig:qxpx_5e7_E4} at large $\xi$ ($\bar{x} \simeq 2\,000\rightarrow 3\,000$ in that figure). Such a population does not arise in a one-dimensional geometry (as we have checked), in which the CFI does not develop.

\section{Exploring different configurations}
\label{sec:varparam}

\subsection{Influence of the ion mass}
\label{subsec:immobile_ions}

\begin{figure}
    \centering
    \includegraphics[width=0.45\textwidth]{ 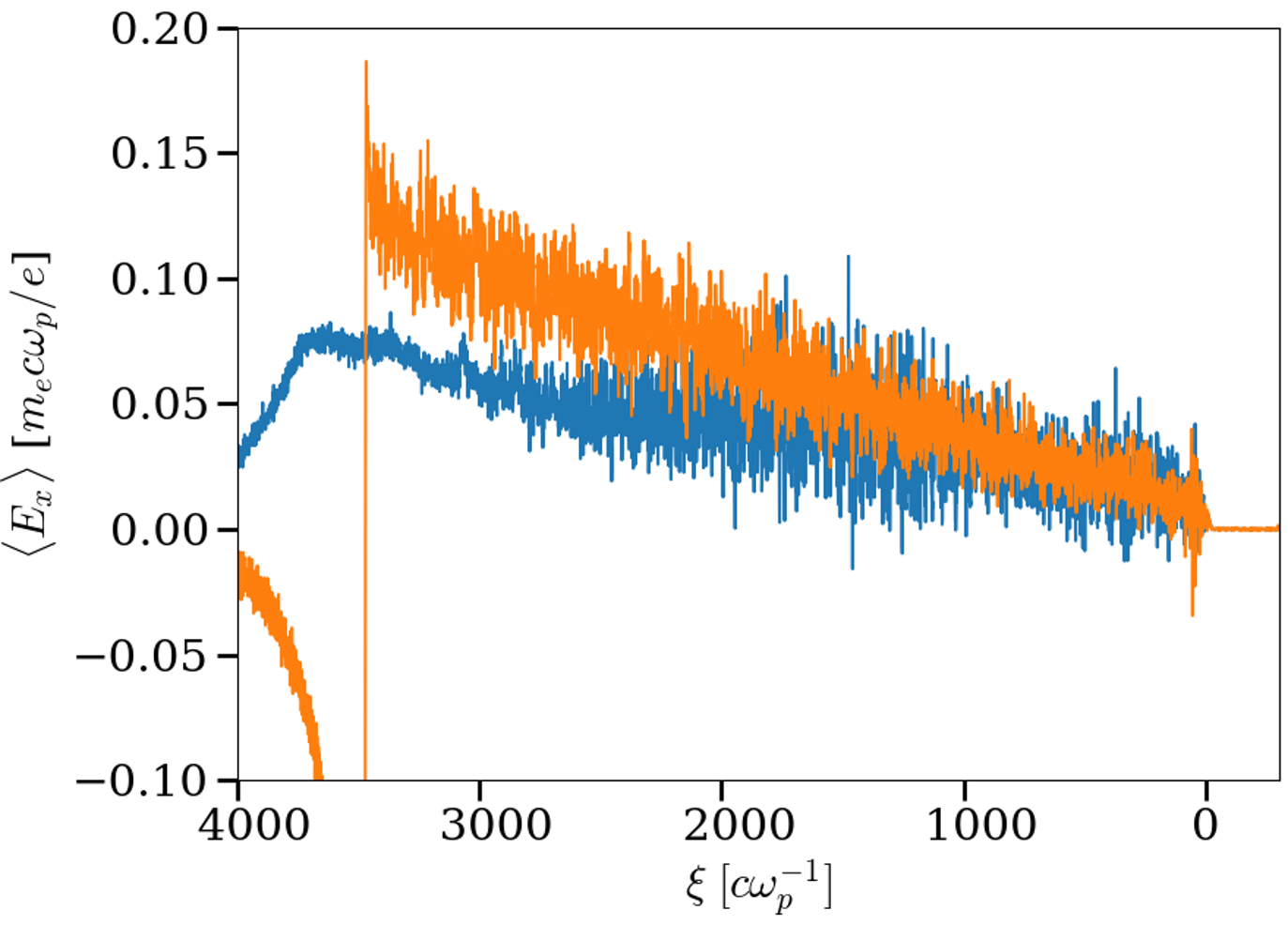}
    \caption{Charge-separation electric field, $\langle E_x \rangle$, in the (blue) $m_i/m_e = 184$ and (orange) $m_i = +\infty$ cases, averaged over the $y$ direction. The other main parameters are $\bar{\epsilon}_\gamma = 4$ and $\alpha_\gamma = 1.05\times 10^4$. Near the photon front, the two longitudinal profiles of $\langle E_x \rangle$ almost coincide, while further away $\langle E_x \rangle$ increases more slowly with distance when $m_i/m_e=184$.}
    \label{fig:Ex_mi_inf}
\end{figure}

\begin{figure}
    \centering
    \includegraphics[width=0.45\textwidth]{ 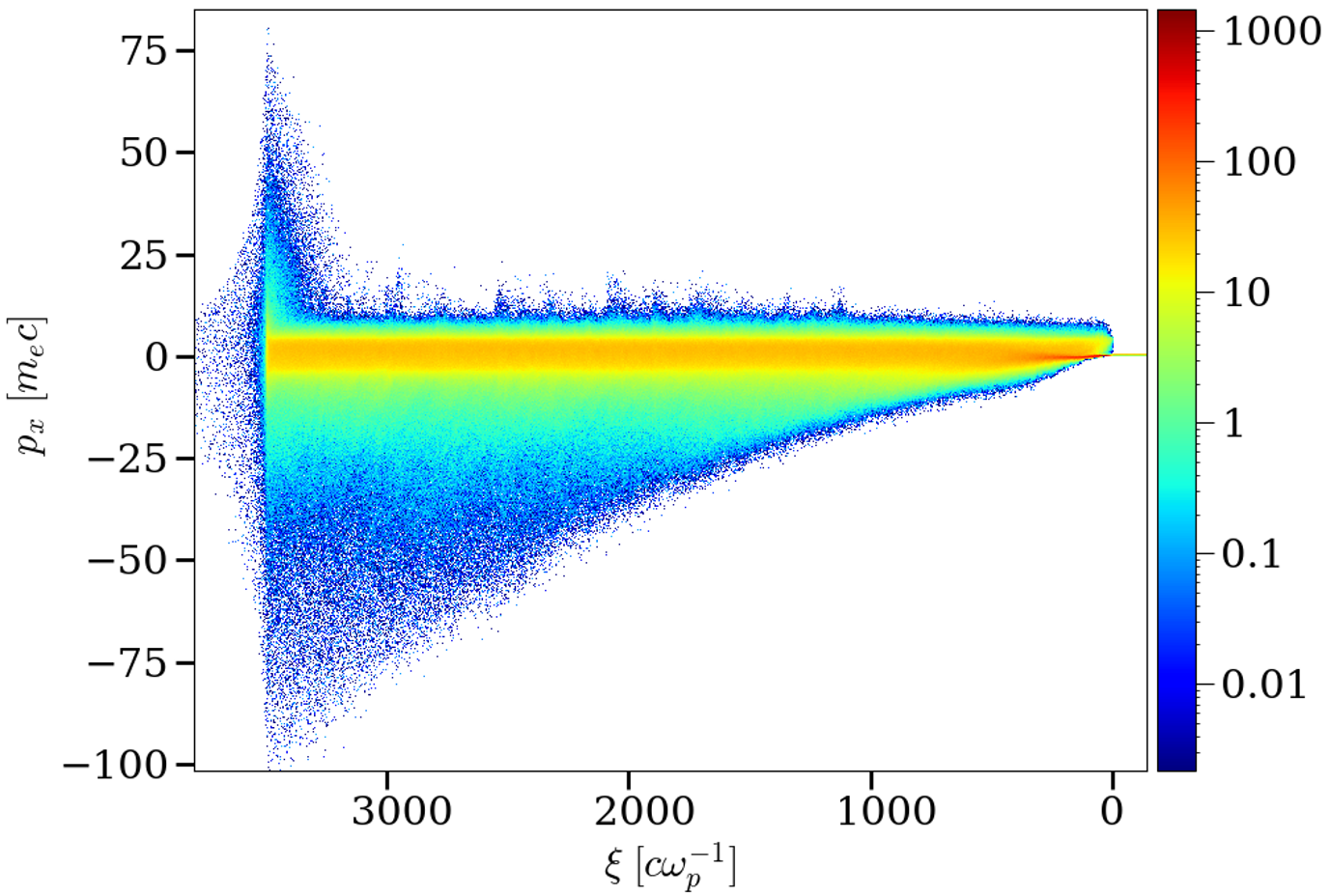}
    \caption{Electron $\xi-p_x$ phase space, using immobile ions along with $\bar{\epsilon}_\gamma = 4$ and $\alpha_\gamma =1.05\times 10^4$.}
    \label{fig:qxpx_5e7_mi_inf}
\end{figure}

We first examine a simulation in which the ions are formally ascribed an infinite mass, so that they remain immobile all throughout. In accordance with our earlier discussion, the coherent electric field maintains its linear scaling with $\xi$ all along, as shown in Fig.~\ref{fig:Ex_mi_inf}, where it is compared to the field measured in the $m_i/m_e=184$ case. As $\Delta \gamma_r$ now exclusively depends on $\gamma^{(0)}$, the profile of the energetic tail with $p_x<0$ remains invariant along $\xi$, following Eq.~\eqref{eq:solfx}. This behavior is manifest in the longitudinal phase space of Fig.~\ref{fig:qxpx_5e7_mi_inf}. The tail with large $p_x>0$ that appears at the plasma edge ($\bar{\xi} \simeq 3\,500$) results from energetic electrons having escaped into vacuum and been pulled back by the sheath field they have themselves induced.

Far from the photon front, although the ions remain motionless, the CFI develops to a level comparable to that observed with $m_i/m_e=184$. However, in accordance with the previous discussion, we do not observe a significant bulk motion of the generated magnetic structures in the lab frame. The absence of ion motion also entails a larger charge-separation electric field, which, given that the magnetic field strength remains similar, effectively reduces the fraction of energetic, backward-moving electrons that can be reflected through scattering off the magnetic filaments. Moreover, the absence of ion motion forbids energy-gain interactions through scattering on those filaments. Accordingly, Fig.~\ref{fig:qxpx_5e7_mi_inf} does not reveal a significant suprathermal electron population with $p_x >0$, except for those mentioned above, at the edge of the plasma slab.

Using ions of physical mass ($m_i/m_e=1836$) yields results similar to the immobile-ion case as long as the ion bulk motion remains small compared to the speed of light. This means, in particular, a negligible forward-propagating suprathermal electron population and a distribution of the backward-moving tail that depends mainly on $\gamma$, and weakly on $\xi$. However, if the interaction is sustained long enough, the ions can approach the speed of light, giving rise to a phenomenology similar to that described in our baseline configuration.

Simulating the latter regime with a physical ion mass is computationally demanding because the ion speed depends approximately on $A \bar{\xi}^2$ with $A\,\propto\, m_i^{-1}$ [Eq.~\eqref{eq:ionspeed}]. For the ions to reach a given speed therefore necessitates a $10\times$ larger space and time scale at $m_i/m_e = 1836$ than at $m_i/m_e = 184$. This constraint prompted us to relax the numerical parameters, by doubling the cell size ($\Delta x = \Delta y = 0.4c/\omega_p$) and halving the number of particles per cell (10 per cell and per species).
The photon density was, moreover, kept constant for $\bar{\xi} \ge 4\,800$ in order to achieve a $n_\gamma/n_0$ ratio comparable with that of the $m_i/m_e=184$ case. 

\begin{figure}
    \centering
    \includegraphics[width=0.5\textwidth]{ 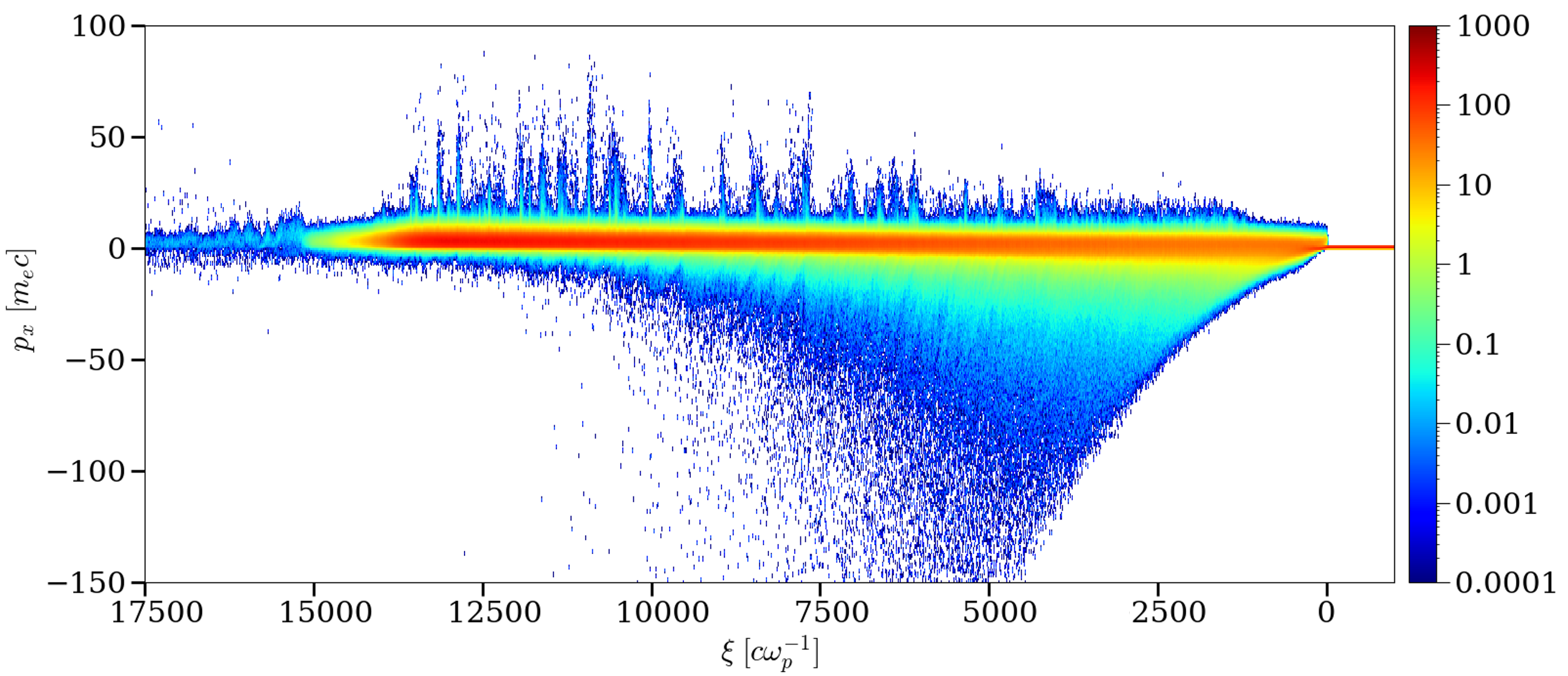}
    \caption{Electron $\xi-p_x$ phase space at $\bar{t} = 28\,000$ in a simulation using $m_i/m_e = 1836$ along with $\bar{\epsilon}_\gamma = 4$ and $\alpha_\gamma =1.05\times 10^4$.}
    \label{fig:qxpx_5e7_mi_phy}
\end{figure}

Figure~\ref{fig:qxpx_5e7_mi_phy} shows the electron $x-p_x$ phase space as extracted at $\bar{t} = 28\,000$ from that large-scale, simulation with $m_i/m_e = 1836$. Both backward- and forward-directed suprathermal electron populations are clearly visible. It is found that the ions reach a drift velocity of $\beta_{ix} \simeq 0.09$ at $\bar{\xi} \simeq 3\,000$, as expected on the basis of Eq.~\eqref{eq:ionspeed}, and $\beta_{ix} \simeq 0.72$ at $\bar{\xi} = 12\,000$. In such conditions, even moderately suprathermal electrons ($\bar{p}_< \simeq -10$) reflecting off the magnetic filaments carried by the ion motion can attain $\bar{p}_> \simeq 80$ (for $\cos\theta_<=-\cos\theta_>=-1$), as observed around $\bar{\xi}=12\,000$.

Interestingly, the backward-directed suprathermal tail appears to be eroded away at distances $\bar{\xi} \gtrsim 3000$.
This follows from the reduction in the charge-separation field, and hence in the $\Delta \gamma_r$ parameter, caused by the increased bulk ion speed [see Eq.~\eqref{eq:Ex_edf_KN_approx}]. 
As an example, at $\bar \xi \simeq 3000$ where $n_\gamma/n_e \simeq 3.2\times 10^7$, one has $\langle E_x \rangle \simeq 0.09\,m_ec\omega_{\rm p}/e$; setting $\gamma^{(0)}=4$ and $\beta_x^{(0)} =-1$ then yields $\Delta \gamma_r \simeq 2.6$. Further away from the photon front ($\bar \xi \simeq 12\,000$), the charge-separation field is slightly lower, $\langle E_x \rangle \simeq 0.07\,m_ec\omega_{\rm p}/e$, although the photon density has there risen to its maximum density of $n_\gamma/n_e \simeq 5\times 10^7$. We then predict $\Delta \gamma_r \simeq 1.3$, significantly smaller than at $\bar \xi = 3\,000$. The electrons are thus more likely to be scattered back to lower energies, which accounts for the progressive depletion of the $p_x<0$ energetic tail observed at large distances.

\subsection{Larger photon density}
\label{subsec:higher_photon_density}

We have run additional simulations using a radiative flux ten times as dense as previously, \emph{i.e.}, $\alpha_\gamma=1.05 \times 10^5$. The main consequence is to boost the piston effect of the radiative force on the plasma  ions. Otherwise, we observe phenomena similar to those reported at lower photon density, notably the acceleration of electrons to large negative momenta in the charge-separation electric field, the development of the CFI and the reflection of energetic electrons off magnetic filaments. 

\begin{figure}
    \centering
    \includegraphics[width=0.45\textwidth]{ 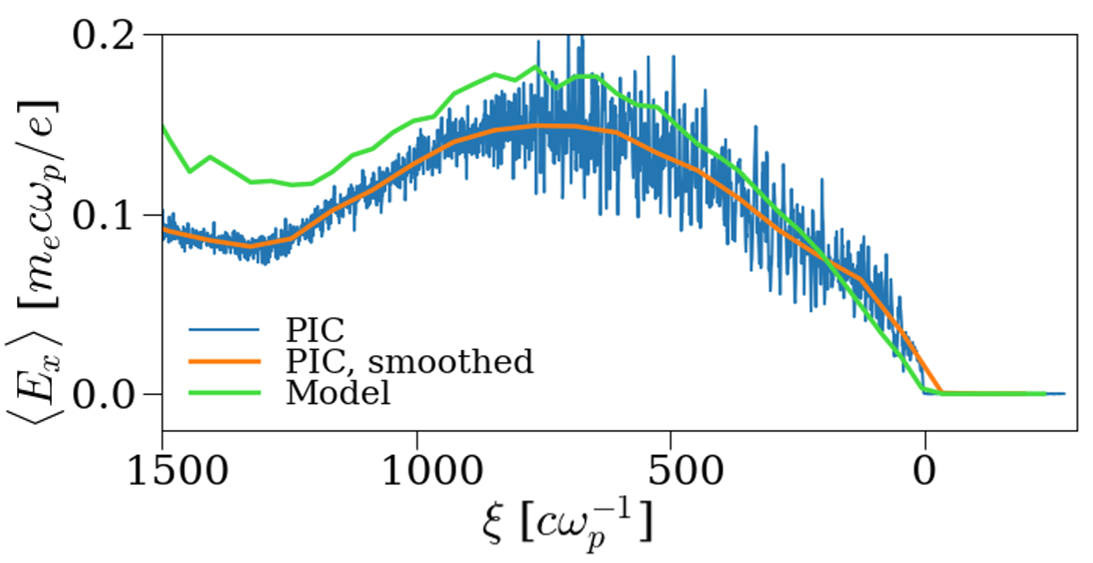}
    \caption{Same as Fig.~\ref{fig:tracethE_5e7} ($m_i/m_e = 184$, $\bar{\epsilon}_\gamma = 4$) but with a larger photon density ($\alpha_\gamma=1.05\times 10^5$).
    }
    \label{fig:tracethE_5e8}
\end{figure}

The charge-separation field in the vicinity of the photon front grows -- expectedly so -- about ten times as fast as for $\alpha_\gamma =1.05\times 10^4$ (see Fig.~\ref{fig:tracethE_5e8}). As a result, the ions are accelerated to relativistic speeds within a few hundreds of $\omega_p^{-1}$; in detail, the ions reach $\bar{p}_{ix} \simeq 2$ at $\bar \xi \simeq 1\,200$, corresponding to a longitudinal velocity $\beta_{ix} \simeq 0.9$. This is accompanied by a substantial compression of the plasma near its boundary, with $n_e \simeq 8 n_0$ being achieved by $\bar{t} \simeq 3\,000$; in comparison, the electron density reaches a maximum of $\sim 2 n_0$ in the reference $\alpha_\gamma = 1.05\times 10^4$ case.

Similarly to the simulation with $m_i/m_e=1836$, the faster bulk plasma motion brings about a significant drop in the charge-separation field beyond $\bar{\xi} \simeq 700$, despite the monotonically rising photon density. The green curve in Fig.~\ref{fig:tracethE_5e8} plots the theoretical prediction of $\langle E_x \rangle$ obtained using a spatially dependent, bi-Maxwellian fit of the momentum electron distribution, as done in Sec.~\ref{subsec:charge_separation}. Again, good agreement with the PIC results is found. To demonstrate the mitigating effect of the plasma motion on $\langle E_x \rangle$, we performed a similar calculation but neglecting the mean drift velocity (\emph{i.e.}, taking $\bar{p}_d=0$) in the bi-Maxwellian distribution. This yielded a curve (not shown) continually growing with $\bar{\xi}$, failing to capture the maximum electric field observed in the simulation.

\begin{figure}
    \centering
    \includegraphics[width=0.45\textwidth]{ 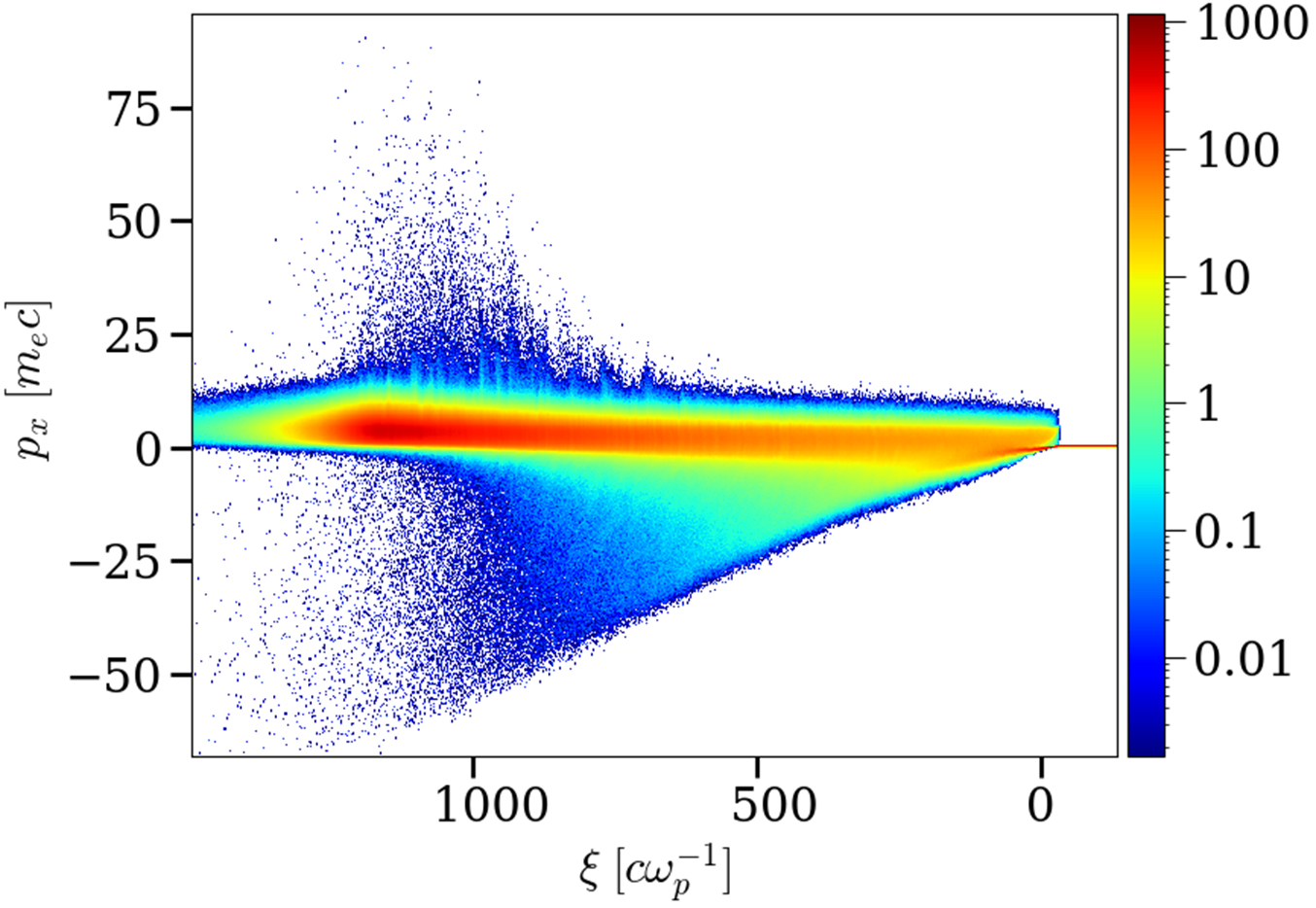}
    \caption{Electron $\xi-p_x$ phase space. The simulation parameters are those of Fig.~\ref{fig:tracethE_5e7} except for a larger photon density ($\alpha_\gamma=1.05\times 10^5$).}
    \label{fig:qxpx_5e8}
\end{figure}

Those larger ion momenta, in particular their relativistic character, have interesting consequences regarding the generation of the suprathermal electron population. For one, the energy attained through one reflection is significantly higher for a same initial momentum. To see this, it proves useful to rewrite Eq.~\eqref{eq:stochastic} as $p_> = \gamma_i^2 \left(1-\beta_{ix} \cos\theta_<\right)\left(1+\beta_{ix} \cos\theta^*_>\right)$ where $\cos\theta^*_>$ is evaluated in the filament frame (indicated with a $^*$ symbol), moving at velocity $\beta_{ix}$ with corresponding Lorentz factor $\gamma_i$. The presence of $\gamma_i^2$ in that equation significantly boosts the energy gain with respect to that observed in the sub-relativistic ($\beta_{ix} \ll 1$) limit.

\begin{figure}
    \centering
    \includegraphics[width=0.4\textwidth]{ 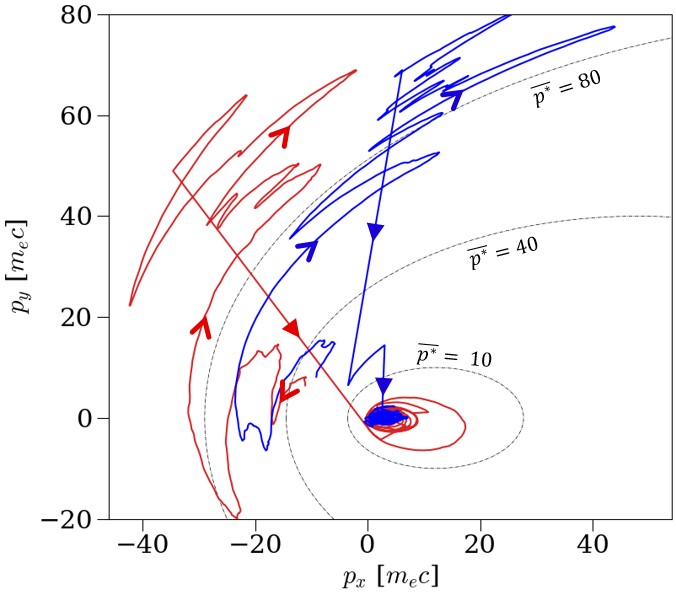}
    \caption{Two representative electron $p_x-p_y$ trajectories from the simulation with $\alpha_\gamma = 1.05\times 10^5$, $\bar{\epsilon}_\gamma = 4$ and $m_i/m_e=184$. As the ions, and therefore the microturbulence, move at relativistic bulk velocities, the loci of constant total momentum in the filament frame ($\bar{p}^*$) are no longer circles, but ellipses. These are indicated in dotted lines, assuming that the filament frame moves at $\beta_{ix} = 0.77$, as measured for ions around $\bar{\xi} \simeq 1\,000$.}
    \label{fig:pxpy_5e8_reflection}
\end{figure}

Figure~\ref{fig:pxpy_5e8_reflection} plots the $p_x-p_y$ trajectories of two macro-electrons in that high-$n_\gamma$ case. To emphasize the influence of relativistic effects, we overlay in dotted lines the loci of constant total momenta ($\bar{p}^*$) as measured at $\bar \xi \simeq 1\,000$ in the filament frame of approximate velocity $\beta_{ix} \simeq 0.77$. The trajectory of the red macro-electron is similar to those seen before. This electron gains large $p_x<0$ momentum while rotating in the $p_x-p_y$ plane under the influence of the CFI-driven $B_z$ fields. Eventually, it loses most of its energy through inverse Compton scattering and goes back to the plasma bulk. The blue electron, on the other hand, is being energized with $p_x>0$ momentum while being scattered. This effect results from the relativistic motion of the magnetic filaments: the longitudinal electron momentum in the filament frame can be written as $p_x^* = \gamma_i p_x \left(1 - \beta_{ix}/\beta_x \right)$.
Obviously, if $0 <\beta_x < \beta_{ix}$, the longitudinal momentum $p_x^*$ is negative even though $p_x>0$. Consequently, in conjunction with angular scattering, the charge-separation electric field acts so as to increase the total momentum $p^*$ as measured in the filament frame. For that particular blue macro-electron, that interplay between the coherent electric field and the moving magnetic filaments leads to an increase in momentum from $\bar{p} \simeq 40$ to $\bar{p} \gtrsim 80$ before it is Compton-scattered back to lower energies. 

\subsection{Finite-length radiative pulse}
\label{subsec:finite_pulse}

\begin{figure}
    \centering
    \includegraphics[width=0.45\textwidth]{ 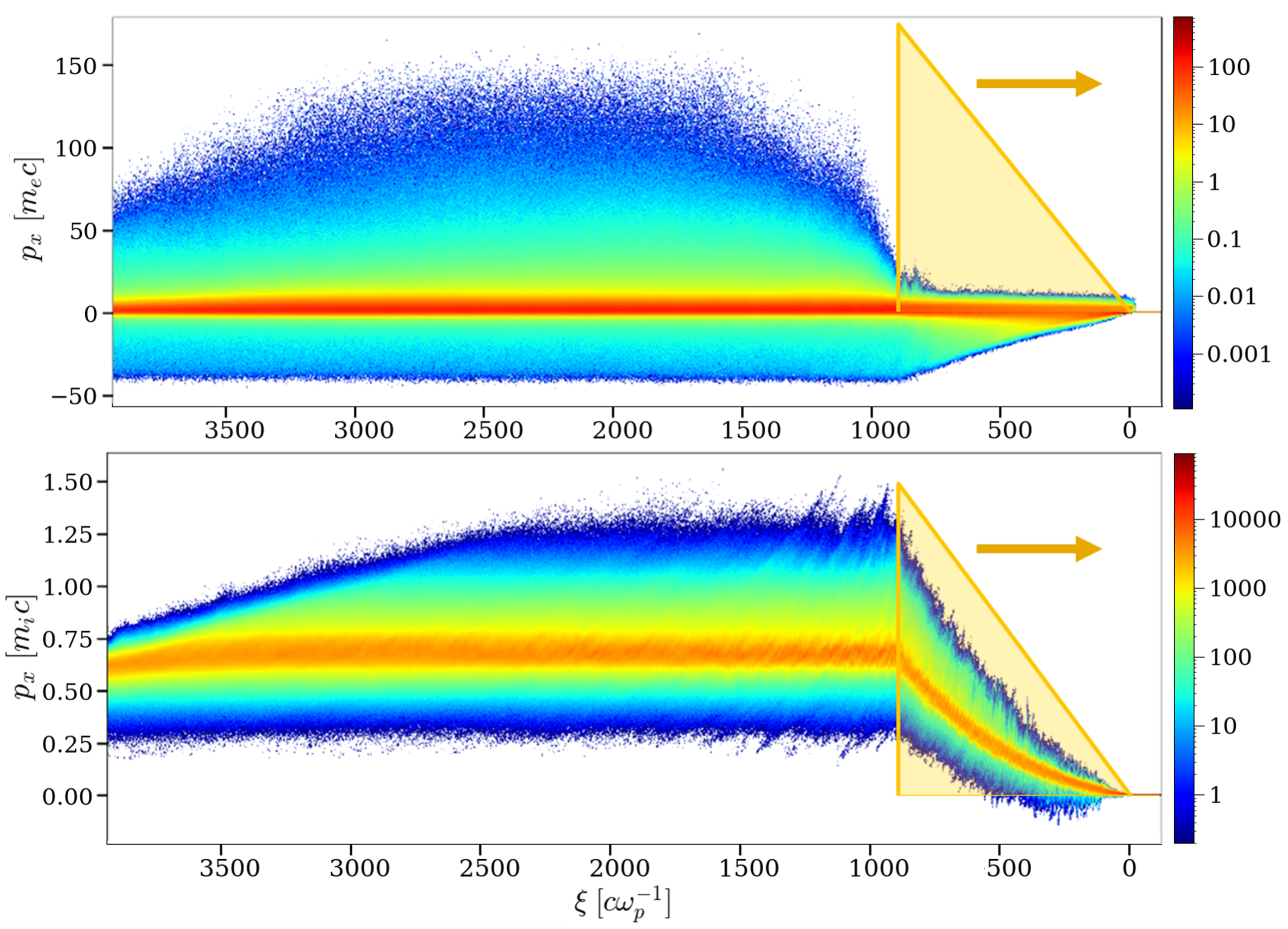}
    \caption{Electron (top) and ion (bottom) $\xi-p_x$ phase spaces at $\bar{t} = 11\,000$, from a simulation using a finite-length photon flux with $\bar{\epsilon}_\gamma = 4$ and $m_i/m_e=184$. The yellow triangles indicate the photon density profile: it increases as $n_\gamma = 1.05\times 10^5\,n_0\,\bar{\xi}$ up to $\bar{\xi}<\bar{\xi}_{\rm max} = 900$, and vanishes further away.
    }
    \label{fig:qxpx_finite_5e8}
\end{figure}

We now explore the consequence of imposing a finite length scale on the photon flux profile. In previous simulations, we observed the formation of a $p_x>0$ suprathermal population through the scattering of electrons with large $p_x<0$ from the CFI-driven magnetic filaments moving at a velocity $\sim \beta_i$. The comparatively small amount of those high-energy electrons with $p_x >0$ resulted from the comparatively large charge-separation field and from the finite probability of undergoing inverse Compton scattering, which both tend to reduce the energy of such electrons and trap them into the plasma bulk. If we now consider a finite-length radiative pulse, the initial radiative force still creates the two previously observed populations, and the CFI develops in the same way as it does with longitudinally infinite radiative fluxes~\cite{APJL_Frederiksen08}. However, in the present case, the charge-separation field vanishes behind the radiative pulse, while the magnetic filaments survive. Consequently, all of the electrons in the energetic $p_x<0$ tail end up being scattered by those filaments, creating a dense population of high-energy electrons with a mean bulk velocity equal to that of the ions.

This feature is clearly illustrated by the electron and ion phase spaces shown in Fig.~\ref{fig:qxpx_finite_5e8}. There, the photon-to-electron density ratio increases as $n_\gamma/n_0 = 1.05\times 10^5\,\bar{\xi}$ up to $\bar{\xi}<\bar{\xi}_{\rm max} = 900$ beyond which $n_\gamma$ drops to zero. One can see that the ion momentum distribution (bottom) becomes uniform along $x$ when the radiative force vanishes, thus evidencing the absence of a coherent electric field, while the electrons (top) are suddenly left free of the influence of Compton scattering and the longitudinal electric field, whence the quick formation of the high-energy population for $\bar{\xi} \gtrsim 1\,000$.

\subsection{Different photon energies}
\label{subsec:photon_energies}

Finally, we investigate the influence of the photon energy $\epsilon_\gamma$. Using lower-energy photons ($\bar \epsilon_\gamma=0.1$) while keeping the same photon density profile with $\alpha_\gamma = 1.05\times 10^4$ (hence effectively reducing the photon energy flux) reduces substantially the radiative force that they exert on the plasma. In particular, the ions cannot reach relativistic velocities during the timespan of the simulation, while the electrons are for the most part scattered before they can be accelerated in significant amounts by the charge-separation field. Compton scattering then indeed takes place in the Thomson regime with, correspondingly, a relatively large cross-section that impedes the formation of a counterpropagating energetic tail.

\begin{figure}
    \centering
    \includegraphics[width=0.45\textwidth]{ 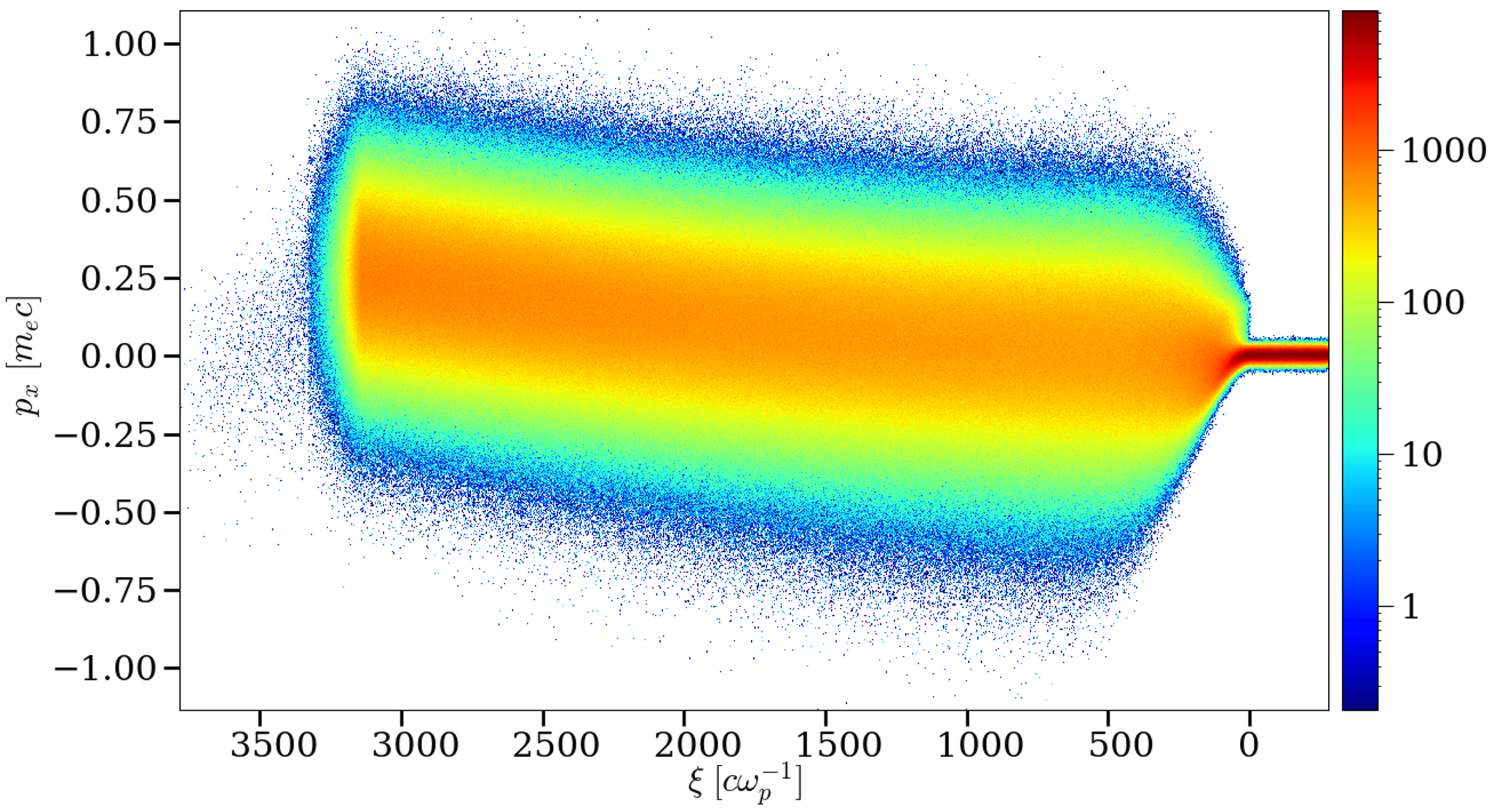}
    \caption{Electron $\xi-p_x$ phase space at $\bar{t} = 3\,500$, from a simulation with  $\bar{\epsilon}_\gamma = 0.1$, $\alpha_\gamma=1.05\times 10^4$ and $m_i/m_e = 184$.}
    \label{fig:qxpx_var_epsG}
\end{figure}

In terms of the $\Delta \gamma_r$ parameter introduced earlier, we have, for an electron momentum $\bar{p}=-\bar{\epsilon}_\gamma^{1/2} = -0.32$ (as discussed in Sec.~\ref{subsec:suprathermal_tail}) 
$\Delta \gamma_r \sim 0.02 \ll \bar{\epsilon}_\gamma$, so that no energetic electron tail is expected to form, and indeed none is seen in the electron phase space reported in Fig.~\ref{fig:qxpx_var_epsG}.  The resulting electron momentum anisotropy still excites filamentation instabilities, the main role of which, however, is to isotropize the various populations in the filament (ion) frame.

At photon energies well in excess of $m_ec^2$, radiative processes that were neglected in the fiducial simulation may become important, in particular those leading to pair production. Regarding Bethe-Heitler pair production~\cite{APJ_Lightman82, MNRAS_Stepney83}, the cross-sections remain much smaller than the Compton cross-section, both in photon-proton interaction, and in photon-electron (triplet) interaction as long as the photon energy in the particle frame lies below $256\,m_ec^2$ (corresponding to an electron Lorentz factor of $\gamma = 32$ for the worst case of an electron propagating in the directly opposite direction to photons of energy $4\,m_ec^2$). We have checked that those processes can be neglected in the present conditions of interest.

Pair production can also take place through direct Breit-Wheeler (photon-photon) interactions, and this process can become quite competitive at large values of $n_\gamma$ and $n_e$. For the present configuration of a unidirectional photon beam, such interactions mainly take place between a beam photon and a photon that has been deflected by an angle $\theta>0$ via Compton scattering off the plasma electrons. Over the plasma length scale $L$, the mean differential number density of such stray photons can be approximated as ${\rm d}n_\gamma'/{\rm d}\mu\simeq \langle n_\gamma\rangle  n_0 L\,{\rm d} \sigma_{\rm C}/{\rm d}\mu$ with $\mu = \cos\theta$, ${\rm d} \sigma_{\rm C}/{\rm d}\mu$ the differential Compton cross-section, $n_e$ the electron plasma density and $\langle n_\gamma\rangle = L^{-1}\int_0^L {\rm d}\xi\, n_\gamma$. The mean number density of pairs created through Breit-Wheeler interactions over the plasma length scale can similarly be estimated as
\begin{equation}
    n_\pm \simeq \int_{-1}^{+1}{\rm d}\mu\, \frac{{\rm d}n_\gamma'}{{\rm d}\mu}\, \sigma_{\rm BW}(s_{\rm BW})\,(1-\mu)\,\langle n_\gamma\rangle\,L\, ,
    \label{eq:BWpp}
\end{equation}
where the Breit-Wheeler cross-section $\sigma_{\rm BW}$~\cite{PhysRev.155.1404} is written in terms of the normalized invariant $s_{\rm BW} \equiv \bar{\epsilon}_\gamma \bar{\epsilon}_\gamma'(1-\mu)/2$, where $\bar{\epsilon}_\gamma$ denotes the beam photon energy and $\bar\epsilon_\gamma' \equiv \bar{\epsilon}_\gamma/\left[1 + \bar\epsilon_\gamma (1 - \mu)\right]$ the energy of the Compton-deflected photon. The back-of-the-envelope estimate $n_{\pm} \sim \mathcal O(\langle n_\gamma\rangle^2 n_e \sigma_{\rm T}^2 L^2)$, which disregards any Klein-Nishina suppression and pair-production threshold effect, gives $n_{\pm}/n_e \sim 10^6$ for the reference simulation with $\bar{\epsilon}_\gamma = 4$. A careful numerical calculation rather predicts $n_{\pm}/n_e \sim 3 \times 10^4$, which still remains much larger than unity.

In the limit $n_{\pm}/n_e > m_i/m_e$, Breit-Wheeler interactions can thus load the plasma with sufficiently many pairs to overcome the ion inertia. The global interaction would then mimic that with a pair plasma, at a photon-to-electron density ratio effectively reduced by $n_{\pm}/n_e$. On the other hand, in physical conditions such that $n_{\pm}/n_e < m_i/m_e$, the bulk plasma dynamics should approximately correspond to that studied here, meaning with an effective mass ratio $(m_i/m_e)\,n_e/(n_e+n_\pm)$. However, in both cases, the charge difference between positrons and electrons can introduce additional charge separation phenomena, such as recently discussed in the context of radiation-mediated shock waves~\cite{MNRAS_Vanthieghem22}.

Of course, pair production operates above a critical photon energy, which can be numerically estimated as  $\bar{\epsilon}_{\gamma, \rm cr} \simeq 2.4$. For $\bar{\epsilon}_\gamma < \bar{\epsilon}_{\gamma, \rm cr}$, the center-of-mass energy $\sqrt{4s_{\rm BW}}$ falls below the pair production threshold, and pairs do not appear. At $\bar{\epsilon}_\gamma \gtrsim \bar{\epsilon}_{\gamma, \rm cr}$, the number density of pairs rapidly reaches its maximum value estimated above, while at larger values of $\bar{\epsilon}_\gamma$, it decreases roughly as $1/\bar{\epsilon}_\gamma^2$, following the combined scalings of the individual cross-sections.

\begin{figure}
    \centering
    \includegraphics[width=0.45\textwidth]{ 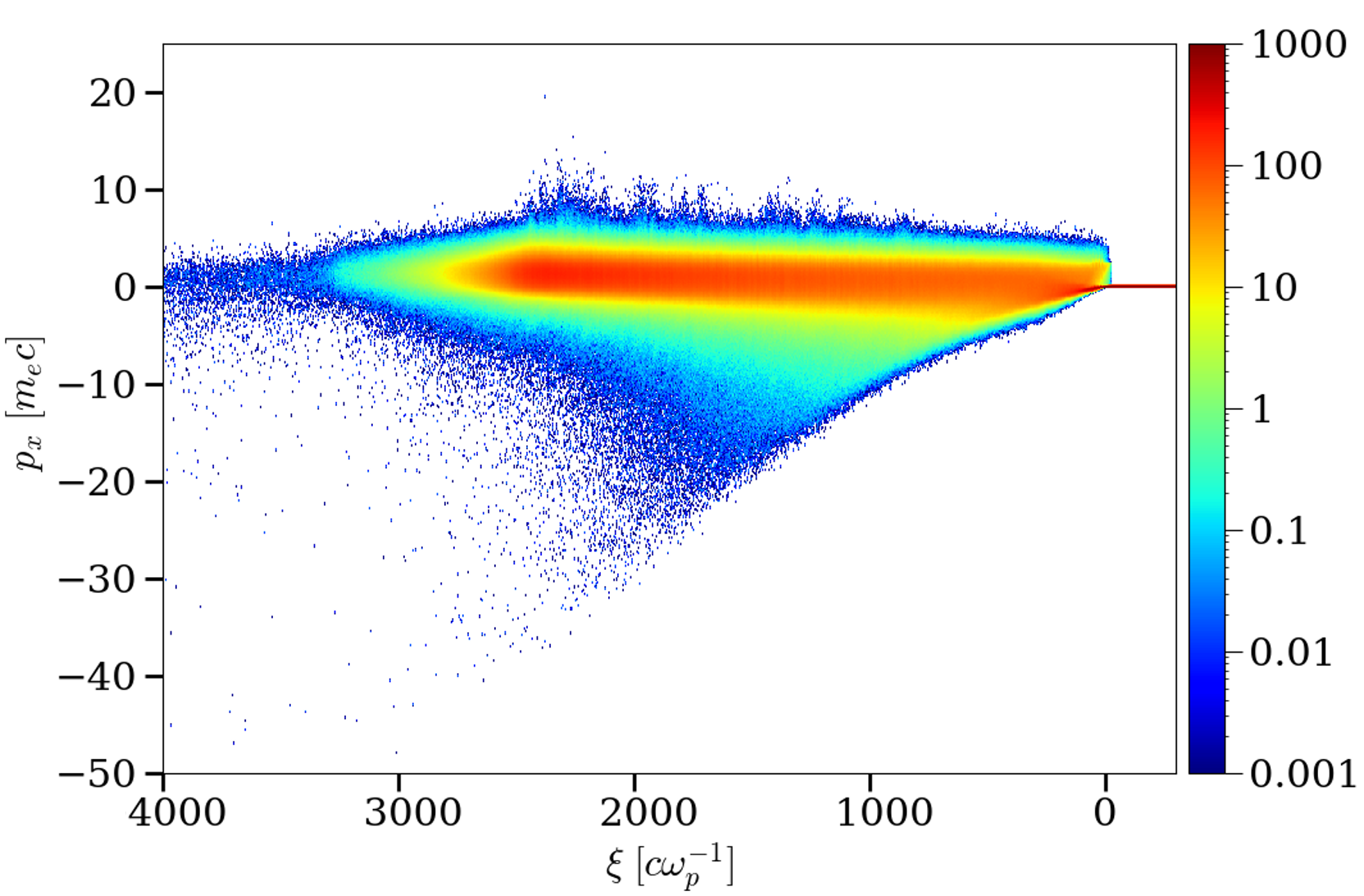}
    \caption{Electron $\xi-p_x$ phase space, from a simulation with $\bar{\epsilon}_\gamma = 2$ and $\alpha_\gamma =1.05\times 10^4$. Features similar to the reference configuration are found, yet in a parameter range precluding photon-photon pair production. }
    \label{fig:qxpx_5e7_E2}
\end{figure}

The relevance of that pair production process has become manifest at a late stage of the present work. Even though our fiducial simulation considers photon energies slightly above threshold, the omission of pair production does not invalidate our results per se. Indeed, the present simulation can be regarded as illustrating the physics of interaction immediately below threshold. To comfort this point of view, we present in Fig.~\ref{fig:qxpx_5e7_E2} the electron phase space from a sub-threshold simulation with $\bar{\epsilon}_\gamma = 2$, all other parameters being equal. It indeed reproduces the features of its counterpart Fig.~\ref{fig:qxpx_5e7_E4}, which formed the basis of the discussion in Sec.~\ref{sec:physcompton}. A realistic simulation for photon energies significantly above threshold, which would properly include the relevant pair production processes, obviously requires substantial algorithmic and computational efforts, and we defer it to future work. Additionally, we stress that the total number density of pairs produced depends sensitively on the physical conditions at hand, since $n_\pm \propto n_\gamma^2 n_e L^2$, so that it make take very different values when the parameters $n_\gamma$, $n_e$ and $L$ are scaled to astrophysical scales of interest while keeping $n_\gamma/\sqrt{n_e}$ constant (as discussed in Sec.~\ref{sec:introduction}). The present simulation can thus also be regarded as simulating the phenomena that would be observed at values of $n_\gamma$ or $n_e$ sufficiently low to avoid excessive pair production.

\section{Conclusions}\label{sec:concl}
\label{sec:conclusions}

We have examined the physics of the interaction between an extremely dense photon flux and a fully ionized, collisionless electron-proton plasma, in the regime where electron-photon Compton scattering takes place in the Klein-Nishina limit. To this effect, we have performed 2D3V PIC simulations using, as reference parameters, a plasma density of $n_e \simeq 10^{21}\,\rm cm^{-3}$, a photon-to-plasma density ratio of $n_\gamma/n_e \simeq 10^7$, a photon energy of $\epsilon_\gamma = 4\,m_ec^2$ and an ion-to-electron mass ratio of $m_i/m_e = 184$. Our numerical experiments have uncovered a complex sequence of particle acceleration processes.

Through Compton interactions, the photons generate a relativistic electron beam that trails immediately behind the photon flux. The slower ion dynamics then entails a charge-separation field $\langle E_x \rangle$ oriented along the radiative flux. An equilibrium between the force exerted by this coherent longitudinal electric field and the radiative force on the electrons establishes rapidly, from which we derive an analytical formula for $\langle E_x \rangle$ that agrees well with the simulation results.

Near the front of the photon beam, two electron populations mix and thereby seed electromagnetic micro-instabilities: the Compton forward-scattered component, and the plasma electrons, which form a return current. The magnetic filaments generated by the current filamentation instability (CFI) are able to deflect the Compton-scattered electrons and inject them into a suprathermal, backward-propagating tail. This population, as well as the non-scattered electron bulk, is progressively accelerated by the charge-separation field to energies largely exceeding those of the driving photons. Those electrons display a runaway-type behavior, as the decrease in the Compton scattering cross-section with energy, in the Klein-Nishina limit, implies that the radiative equilibrium of forces is lost to the benefit of the longitudinal electric field. We provide an analytical approximation to the energy distribution of that suprathermal tail and successfully compare it with the simulation data.

Further away from the photon front, the CFI is sustained by the continuous Compton forward acceleration of a fraction of the electrons, forming an electron component streaming with respect to the bulk.
The latter population is made up of electrons with energy comparable with, or smaller than that of the photons. The interplay of scattering from the CFI fields, of the longitudinal electric field and of occasional (inverse) Compton scattering events acts to maintain those electrons inside the bulk.

The charge-separation field also forward-accelerates the plasma ions, which, as we describe with a simple fluid model, can attain near-relativistic speeds after a $\sim 1\,000\,\omega_{\rm p}^{-1}$ interaction time ($\sim 10\,000\,\omega_{\rm p}^{-1}$ for a physical ion mass $m_i/m_e = 1836$). Due to current balance, this causes the electrons, and hence the magnetic filaments, to acquire the same mean velocity. Upon reflecting off the fast-moving filaments in a Fermi-type fashion, part of the energetic, backward-propagating electrons can be accelerated to form a higher-energy, forward-propagating electron tail.

The simplicity of the driving phenomenon -- a dense gamma-ray burst -- makes the processes described in this work applicable to a broad variety of extreme astrophysical scenarios or in the relativistic laboratory astrophysics experiments envisioned in the future. Also, the suprathermal tails generated in the above photon-plasma interactions may have interesting phenomenological signatures through secondary radiative processes (e.g. synchrotron radiation) which have been neglected in the present study.
Finally, the possibility of pair production via interactions between incoming and Compton-scattered photons deserves a separate investigation. To a first approximation, pair production would act to reduce the average ion-to-electron mass ratio, and thus enhance the phenomenological signatures discussed above. Those effects will be included in a future study. 

\begin{acknowledgments}
This project was provided with computer and storage resources by GENCI at
TGCC thanks to the grant 2022-A0130512993 on the Joliot Curie supercomputer's ROME and SKL partitions.
This work is supported by the French National Research Agency (ANR UnRIP project, Grant No.~ANR-20-CE30-0030). We thank X.~Davoine for his assistance with the \textsc{calder} code.
\end{acknowledgments}

\appendix

\section{Analytical approximation of the charge-separation field}
\label{app:Ex}

We have not managed to obtain an analytical estimate of $\langle E_x \rangle$ in the Klein-Nishina limit ($\eta \gg 1$) using the above bi-Maxwellian fit to $f_e(\mathbf{p})$. Yet a useful formula can be derived in the case of a drifting relativistic thermalized plasma, $f_e(\mathbf{\bar{p}}) = \frac{\mu'^2 n_e}{2\pi \gamma_d} e^{-\mu'\gamma_d (\gamma -\beta_d \bar{p}_x)}$, where $\beta_d$ is the longitudinal bulk speed, $\gamma_d \equiv 1/\sqrt{1-\beta_d^2}$ and $1/\mu' \equiv \widebar{T}' \equiv \gamma_d \langle \bar{p}_y^2/\gamma \rangle \gg 1$ is the normalized proper electron temperature (a $p_x-p_y$ momentum space is considered).
Making use of the large-$\eta$ expansion $G(\eta) \simeq \frac{3}{8\eta}(\ln 2\eta -5/6)$ \cite{ApJ_Blumenthal74}, the electric field can be first approximated to
\begin{align}
    e\langle E_x \rangle &\simeq \frac{3 \sigma_{\rm T} m_{\rm e} c^2 n_\gamma \mu'^2}{16\pi \gamma_d} \int \frac{d^2 \bar{p}}{\gamma} e^{-\mu' \gamma_d (\gamma -\beta_d \bar{p}_x)} \nonumber \\
    & \times \left\{ \ln \left[ 2\bar{\epsilon}_\gamma (\gamma - \bar{p}_x)\right] - \frac{5}{6} \right\} \left(1-\frac{\bar{p}_x}{\bar{\epsilon}_\gamma} \right) \,. 
    \label{eq:Ex_edf_KN_approx_1}
\end{align}
We then change to the integration variables $\bar{p}_x' \equiv \gamma_d(\bar{p}_x-\beta_d \gamma)$ and $\bar{p}_y' \equiv \bar{p}_y$, so that
\begin{align}
    e\langle E_x \rangle &\simeq \frac{3 \sigma_{\rm T} m_{\rm e} c^2 n_\gamma \mu'^2}{16\pi \gamma_d} \int \frac{d^2 p'}{\gamma'} e^{-\mu' \gamma'} \nonumber \\
    & \times \left\{ \ln \left[ 2\bar{\epsilon}_\gamma \sqrt{\frac{1-\beta_d}{1+\beta_d}} \left(\gamma'-\bar{p}_x' \right) \right] - \frac{5}{6} \right\} \nonumber \\
    &\times \left[1-\frac{\gamma_d}{\bar{\epsilon}_\gamma} \left(\bar{p}_x'+\beta_d \gamma' \right) \right] \,. \label{eq:Ex_edf_KN_approx_2}
\end{align}
We next consider polar coordinates, $d^2\bar{p}'\equiv \bar{p}' d\bar{p}' d\theta'$, and assume $\bar{p}' \simeq \gamma' \gg 1$. This leads to
\be
    e\langle E_x \rangle \simeq \frac{3 \sigma_{\rm T} m_{\rm e} c^2 n_\gamma \mu'^2}{16\pi \gamma_d} \int_1^\infty d\gamma' e^{-\mu' \gamma'}
    \int_0^{2\pi} d\theta' I(\gamma',\theta') \,,
    \label{eq:Ex_edf_KN_approx_3}
\ee
where we have introduced
\begin{align}
    I(\gamma',\theta') &=
    \left\{ \ln \left[ 4 \bar{\epsilon} _\gamma \sqrt{\frac{1-\beta_d}{1+\beta_d}} \gamma' \sin^2(\theta'/2) \right] - \frac{5}{6} \right\} \nonumber \\
    &\times \left[1- \frac{\gamma_d}{\bar{\epsilon}_\gamma} \gamma' \left(\cos \theta' + \beta_d \right) \right] \,.
\end{align}
The integration over $\theta'$ can be exactly performed:
\begin{align}
    &\int_0^{2\pi} d\theta'\,I(\gamma',\theta') = 2\pi \Bigg\{ \left(1- \frac{\gamma_d \beta_d}{\bar{\epsilon}_\gamma} \gamma' \right) \nonumber \\
    & \times \left[ \ln \left( \bar{\epsilon}_\gamma \sqrt{\frac{1-\beta_d}{1+\beta_d} \gamma'} \right) - \frac{5}{6} \right]
    + \frac{\gamma_d'}{\bar{\epsilon}_\gamma} \gamma' \Bigg\} \,.   
\end{align}
Inserting this expression into Eq.~\eqref{eq:Ex_edf_KN_approx_3} and making use of the formula $\int_1^\infty \mathrm{d}\gamma\,e^{-\mu \gamma} \ln \gamma = E_1(\mu)/\mu$ [$E_1(\mu)$ is the exponential integral] and of the large-$\mu$ expansion $E_1(\mu) \sim -\ln \mu-\gamma_{\rm E}$ ($\gamma_{\rm E}=0.5772\ldots$ is the Euler constant), one can obtain after some algebra
\begin{align}
    &e\langle E_x \rangle \simeq \frac{3}{8} \sigma_{\rm T} m_{\rm e}c^2 \frac{n_\gamma}{\bar{\epsilon}_\gamma} \Bigg\{\left( \frac{\mu' \bar{\epsilon}_\gamma}{\gamma_d} - \beta_d \right) \nonumber \\
    & \times \left[ \ln \left(\frac{\bar{\epsilon}_\gamma}{\mu'} \sqrt{\frac{1-\beta_d}{1+\beta_d}} \right) - \gamma_{\rm E} - \frac{5}{6}\right] + 1- \beta_d \Bigg\} \,,
    \label{eq:Ex_edf_KN_approx_4}
\end{align}
to leading order in $\mu' \ll 1$ (relativistically hot plasma) and $\eta \gg 1$ (Klein-Nishina limit). Finally, substituting the spatial profile of the photon flux density leads to the following complete expression for the normalized electric field:
\begin{align}
    &\frac{e \langle E_x \rangle}{m_e c \omega_{\rm p}} \simeq \frac{3}{8} \sigma_{\rm T} \frac{c}{\omega_{\rm p}} \frac{\alpha_\gamma n_0 \bar{\xi}}{\bar{\epsilon}_\gamma} \Bigg\{\left(\frac{\mu' \bar{\epsilon}_\gamma}{\gamma_d} - \beta_d \right) \nonumber \\
    & \times \left[ \ln \left(\frac{\bar{\epsilon}_\gamma}{\mu'} \sqrt{\frac{1-\beta_d}{1+\beta_d}} \right) - \gamma_{\rm E} - \frac{5}{6}\right] + 1- \beta_d \Bigg\} \,.
    \label{eq:Ex_edf_KN_approx_5}
\end{align}

We now address the Thomson limit ($\eta \ll 1$) in the case of low-energy photons ($\bar{\epsilon}_\gamma \ll 1$) photons and nonrelativistic electrons, $\widebar{T} \equiv 1/\mu  = \mathcal{O}(\bar{\epsilon}_\gamma) \ll 1$. Using the small-$\eta$ expansion $G(\eta) \simeq 1-\frac{16}{5}\eta$ \cite{ApJ_Blumenthal74}, assuming negligible bulk motion and keeping the dominant terms in the integrand of Eq.~\eqref{eq:force-eq} results in
\be
\frac{e\langle E_x \rangle}{m_e c \omega_{\rm p}} \simeq \sigma_{\rm T} \frac{c}{\omega_{\rm p}} n_0 \alpha_\gamma \bar{\xi} \bar{\epsilon}_\gamma \left(1 + 2\bar{T} -\frac{11}{5}\bar{\epsilon}_\gamma \right) \,. 
\label{eq:Ex_edf_Thomson_approx_1}
\ee
Note that this estimate differs from that derived in Ref.~\cite{1981ApJ...243L.147O} for an arbitrarily hot plasma but assuming $G(\eta) \sim 1$.

\bibliography{references}

\end{document}